\documentclass[10pt,letterpaper]{article}
\usepackage{opex3,graphicx,amsmath,amssymb, dsfont,subfigure,soul,framed,xcolor}
\colorlet{shadecolor}{yellow} 
\newcommand{\mean}[1]{\mbox{$\langle #1 \rangle$}}
\newcommand{\eff}{\text{eff}}
\newcommand{\opt}{\text{opt}}
\newcommand{\rad}{\text{rad}}
\newcommand{\cmb}{\text{cmb}}
\newcommand{\con}{\text{con}}
\newcommand{\tms}{\text{tms}}
\begin{document}
\title{Suppression of extraneous thermal noise in cavity optomechanics}

\author{Y. Zhao, D. J. Wilson, K.-K. Ni, and H. J. Kimble$^{*}$}

\address{Norman Bridge Laboratory of Physics, 12-33, California Institute of Technology, Pasadena, California 91125}
\email{$^*$hjkimble@caltech.edu}
\pagestyle{plain}

\begin{abstract}
Extraneous thermal motion can limit displacement sensitivity and radiation pressure effects, such as optical cooling, in a cavity-optomechanical system.  Here we present an active noise suppression scheme and its experimental implementation.  The main challenge is to selectively sense and suppress extraneous thermal noise without affecting motion of the oscillator. Our solution is to monitor two modes of the optical cavity, each with different sensitivity to the oscillator's motion but similar sensitivity to the extraneous thermal motion.  This information is used to imprint ``anti-noise'' onto the frequency of the incident laser field. In our system, based on a nano-mechanical membrane coupled to a Fabry-P\'{e}rot cavity, simulation and experiment demonstrate that extraneous thermal noise can be selectively suppressed and that the associated limit on optical cooling can be reduced.
%Extraneous thermal motion can limit displacement sensitivity and radiation pressure effects in a cavity-optomechanical system.  We demonstrate how to suppress this noise by mapping a measurement of the extraneous motion onto the frequency of the incident field.   Our strategy involves monitoring the resonanance frequency of two cavity modes with different sensitivity to the intracavity oscillator's motion but similar sensitivity to extraneous motion.  As an important application, we show how our technique can be used to ``purify" a field used to optically cool the intracavity oscillator.     To demonstrate that the technique works in practice, we have used it to suppress mirror thermal motion in a room temperature ``membrane-in-the-middle" system.   Simulation and experiment demonstrate that extraneous thermal noise can be selectively suppressed and that the associated limit on optical cooling can be reduced.
% 
\end{abstract}


%\maketitle

\section{Introduction}\label{se:Intro}
The field of cavity opto-mechanics \cite{OmTheory:Braginsky1970} has experienced remarkable progress in recent years \cite{OmRev:VahalaOE07, OmRev:Marquardt09, OmRev:Clelend09, OmRev:Aspelmeyer10}, owing much to the integration of micro- and nano-resonator technology. Using a combination of cryogenic pre-cooling \cite{OmExp:AspelmeyerDemo2009, OmExp:KippenbergCryogenic09, OmExp:ClelendControl10} and improved fabrication techniques \cite{OmTheory:PainterModel09,OmExp:AspelmeyerTunneling11, OmExp:HarrisMembrane08,OmExp:KippenbergUltralow08}, it is now possible to realize systems wherein the mechanical frequency of the resonator is larger than both the cavity decay rate and the mechanical re-thermalization rate  \cite{OmExp:KippenbergSidebandCooling11PRA, OmExp:Kimble2009,OmExp:SimmonsSideband11,OmExp:PainterGroundCooling11,HarrisCommunication}.  These represent two basic requirements for ground-state cooling using cavity back-action \cite{OmTheory:KippenbergGroundStateCooling07, OmTheory:AspelmeyerGroundStateCooling08,OmTheory:Marquardt07PRL}, a milestone which has recently been realized in several systems \cite{OmExp:ClelendControl10,OmExp:SimmonsSideband11, OmExp:PainterGroundCooling11}, signaling the emergence of a new field of  cavity ``quantum'' opto-mechanics \cite{OmRev:Aspelmeyer10}.

Reasons why only a few systems have successfully reached the quantum regime \cite{OmExp:ClelendControl10,OmExp:SimmonsSideband11, OmExp:PainterGroundCooling11} relate to additional fundamental as well as technical sources of noise.  Optical absorption, for example, can lead to thermal path length changes giving rise to mechanical instabilities \cite{OmExp:KippenbergCryogenic09, OmFbExp:Vahala05}. In cryogenically pre-cooled systems, absorption can also introduce mechanical dissipation by the excitation of two-level fluctuators \cite{OmExp:KippenbergCryogenic09,OmExp:KippenbergSidebandCooling11PRA}. Both effects depend on the material properties of the resonator. Another common issue is laser frequency noise, which can produce random intra-cavity intensity fluctuations.  The radiation pressure associated with these intensity fluctuations can lead to mechanical heating sufficient to prevent ground state cooling  \cite{Thermal:Saulson90, OmTheory:LWHazard, OmTheory:MeystrePhaseNoise11}.  A fully quantum treatment of laser frequency noise heating in this context was recently given in \cite{OmTheory:RablPhaseNoise09}.

In this paper we address an additional, ubiquitous source of extraneous noise -- thermal motion of the cavity apparatus (including substrate and supports) -- which can dominate in systems operating at room temperature.  
Thermal noise is well understood to pose a fundamental limit on mechanics-based measurements \cite{Thermal:Saulson90} spanning a broad spectrum of applications, including gravitational interferometry \cite{OmTheory:Overlap, Thermal:Harry06}, atomic force microscopy \cite{Thermal:Butt95}, ultra-stable laser reference cavities \cite{Thermal:Numata04}, and NEMS/MEMS based sensing \cite{Thermal:Roukes02, Thermal:Gabrielson93}.  Conventional approaches to its reduction involve the use of low loss construction materials \cite{Thermal:BraginskyBook86, Thermal:Penn06} and cryogenic operation temperatures \cite{OmExp:ClelendControl10,OmExp:SimmonsSideband11, OmExp:PainterGroundCooling11}, as well as various forms of feedback \cite{Thermal:Levin01, OmFbExp:PinardCool99, OmFbExp:RugarCooling07, OmFbExp:McclellandSuppression05,OmExp:Karrai04}.   Indeed, schemes for optomechanical cooling \cite{OmRev:Clelend09,OmTheory:BraginskyTranquilizer02} were developed to address this very problem, with the focus on suppression of thermal noise associated with a single oscillatory mode of the system.

Here we are concerned specifically with \emph{extraneous} thermal motion of the apparatus.  In a cavity optomechanical system,  this corresponds to structural vibrations \emph{other than} the mode under study, which lead to extraneous fluctuations of the cavity resonance frequency. For example, in a ``membrane-in-the-middle'' cavity optomechanical system \cite{OmExp:HarrisMembrane08,OmExp:Kimble2009, OmExp:Harris08Nature,OmExp:HarrisDispersive08}, the extraneous noise is the thermal noise of the cavity mirrors, while the vibrational mode under study is of the membrane. Like laser frequency noise \cite{OmTheory:MeystrePhaseNoise11,OmTheory:RablPhaseNoise09}, these extraneous fluctuations can lead to noise heating as well as limit the precision of displacement measurement.    
To combat this challenge, we here propose and experimentally demonstrate a novel technique to actively suppress extraneous thermal noise 
 in a cavity opto-mechanical system.  A crucial requirement in this setting is the ability to sense and differentiate extraneous noise from intrinsic fluctuations produced by the oscillator's motion.  To accomplish this, our strategy is to monitor the resonance frequency of multiple spatial modes of the cavity, each with different sensitivity to the oscillator's motion but comparable sensitivity to extraneous thermal motion \cite{YeZollerDiscussion}.  We show how this information can be used to electro-optically imprint ``anti-noise'' onto the frequency of the incident laser field, resulting in suppression of noise on the instantaneous cavity-laser detuning.  In the context of our particular system,  based on a nano-mechanical membrane coupled to a Fabry-P\'{e}rot cavity,  simulation and experimental results show that extraneous noise can be substantially suppressed without diminishing back action forces on the oscillator, thus enabling lower optical cooling base temperatures.

\begin{figure}[t]
\centering
\includegraphics[width=11cm]{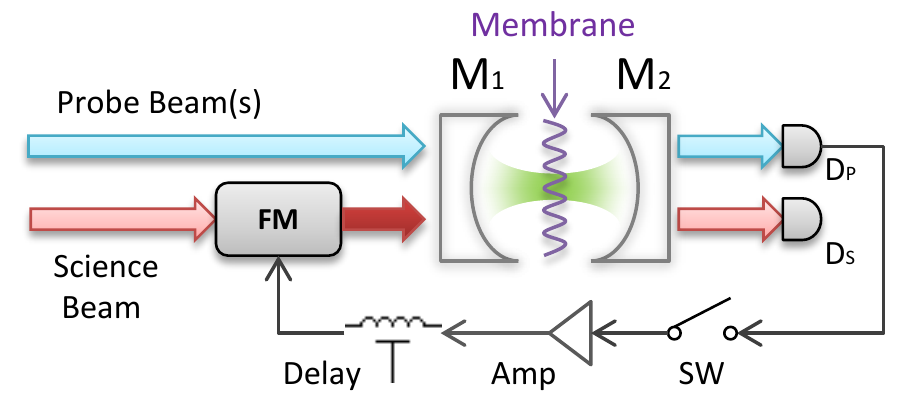}
\caption{Conceptual diagram of the noise suppression scheme. M$_{1/2}$: cavity mirrors. D$_\text{P/S}$: photodetector for the probe field and the science field, respectively. SW: switch. FM: electro-optic frequency modulator.} 
\label{fig:Sch}
\end{figure}

Our paper is organized as follows: in Section \ref{se:Noise} we present an example of extraneous thermal noise in a cavity opto-mechanical system.  In Sections \ref{se:Solution} and \ref{se:Exp} we propose and implement a method to suppress this noise using multiple cavity modes in conjunction with feedback to the laser frequency.  In Section \ref{se:Cooling} we analyze how this feedback affects cavity back-action.  Related issues are discussed in Section \ref{se:Discussion} and a summary is presented in Section \ref{se:Con}. Details relevant to each section are presented in the appendices.

\section{Extraneous thermal noise: illustrative example}\label{se:Noise}

Our experimental system is the same as reported in \cite{OmExp:Kimble2009}. It consists of a high-$Q$ nano-mechanical membrane coupled to a Fabry-P\'{e}rot cavity (Figure \ref{fig:Sch}) with a finesse of $\mathcal{F}\sim10^4$ (using the techniques pioneered in \cite{OmExp:HarrisDispersive08,OmExp:Harris08Nature,OmExp:HarrisStrongCoupling10}).   Owing to the small length ($\mean{L}\simeq 0.74$ mm) and mode waist ($w_{0}\simeq 33\,\mu$m) of our cavity, thermal motion of the end-mirror substrates gives rise to large fluctuations of the cavity resonance frequency, $\nu_{c}$.

\begin{figure}[t]
\centering
\includegraphics[width=13cm]{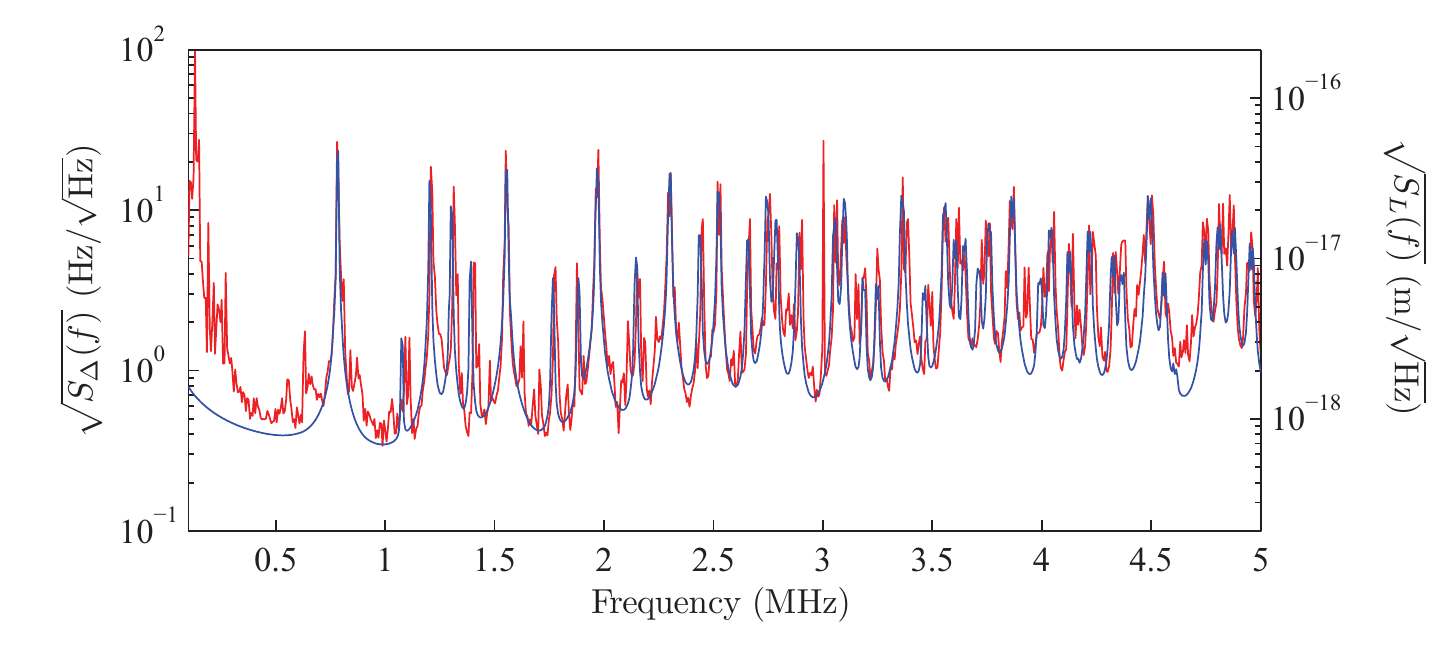}
\caption{Measured spectrum of detuning fluctuations, $\sqrt{S_{\Delta}(f)}$ (also expressed as effective cavity length fluctuations, $\sqrt{S_L(f)}$) for the Fabry-P\'{e}rot cavity described in Section \ref{se:Noise} . The observed noise (red trace) arises from thermal motion of the end-mirror substrates, in agreement with the finite element model (blue trace, Appendix \ref{ap:SubThermal}).  This ``substrate noise'' constitutes an extraneous background for the ``membrane-in-the-middle'' system conceptualized in Figure \ref{fig:Sch} and detailed in \cite{OmExp:Kimble2009}.}
\label{fig:Sub_Mod}
\end{figure}

To measure this ``substrate noise'', we monitor the detuning, $\Delta$, between the cavity (with membrane removed) and a stable input field with frequency $\nu_0 = \nu_{c} +\Delta$.  This can be done using the Pound-Drever-Hall technique \cite{PDH,PDH_Black}, for instance, or by monitoring the power transmitted through the cavity off-resonance ($\langle \Delta \rangle \ne 0$).  A plot of $S_{\Delta}(f)$, the power spectral density of detuning fluctuations \cite{PSD}, is shown in red in Figure \ref{fig:Sub_Mod}.  For illustrative purposes, we also express the noise as ``effective cavity length'' fluctuations $S_{L}(f) = (\mean{L}/\mean{\nu_{c}})^2 S_{\Delta}(f)$. The measured noise between 500 kHz and 5 MHz consists of a dense superposition of $Q\sim 700$ thermal noise peaks at the level of $\sqrt{S_\Delta(f)}\sim 10\text{ Hz}/\sqrt{\text{Hz}}$ ($\sqrt{S_{L}(f)}\sim 10^{-17}\text{ m}/\sqrt{\text{Hz}}$), consistent with the noise predicted from a finite element model of the substrate vibrational modes (Appendix \ref{ap:SubThermal}), shown in blue.  

The light source used for this measurement and all of the following reported in this paper was a Titanium-Sapphire laser (Schwarz Electro-Optics) operating at a wavelength of $\lambda_0=c/\nu_0 \approx 810\text{ nm}$.  In the Fourier domain shown in Figure \ref{fig:Sub_Mod}, an independent measurement of the power spectral density of $\nu_0$ \cite{PSD} gives an upper bound of  $\sqrt{S_{\nu_0}(f)}\le 0.1\text{ Hz}/\sqrt{\text{Hz}}$, suggesting that laser frequency noise is not a major contributor to the inferred $S_\Delta(f)$.

We can gauge the importance of the noise shown in Figure \ref{fig:Sub_Mod} by considering the cavity resonance frequency fluctuations produced by thermal motion of the intra-cavity mechanical oscillator: in our case a 0.5 mm $\times$ 0.5 mm $\times$ 50 nm high-stress ($\approx900$ MPa) Si$_3$N$_4$ membrane with a physical mass of $m_p = 33.6 \text{ ng}$ \cite{OmExp:Kimble2009}.  The magnitude of $S_\Delta(f)$ produced by a single vibrational mode of the membrane depends sensitively on the spatial overlap between the vibrational mode-shape and the intensity profile of the cavity mode (Appendix \ref{ap:Overlap}).

\begin{figure}[t]
\centering
\includegraphics[width=13cm]{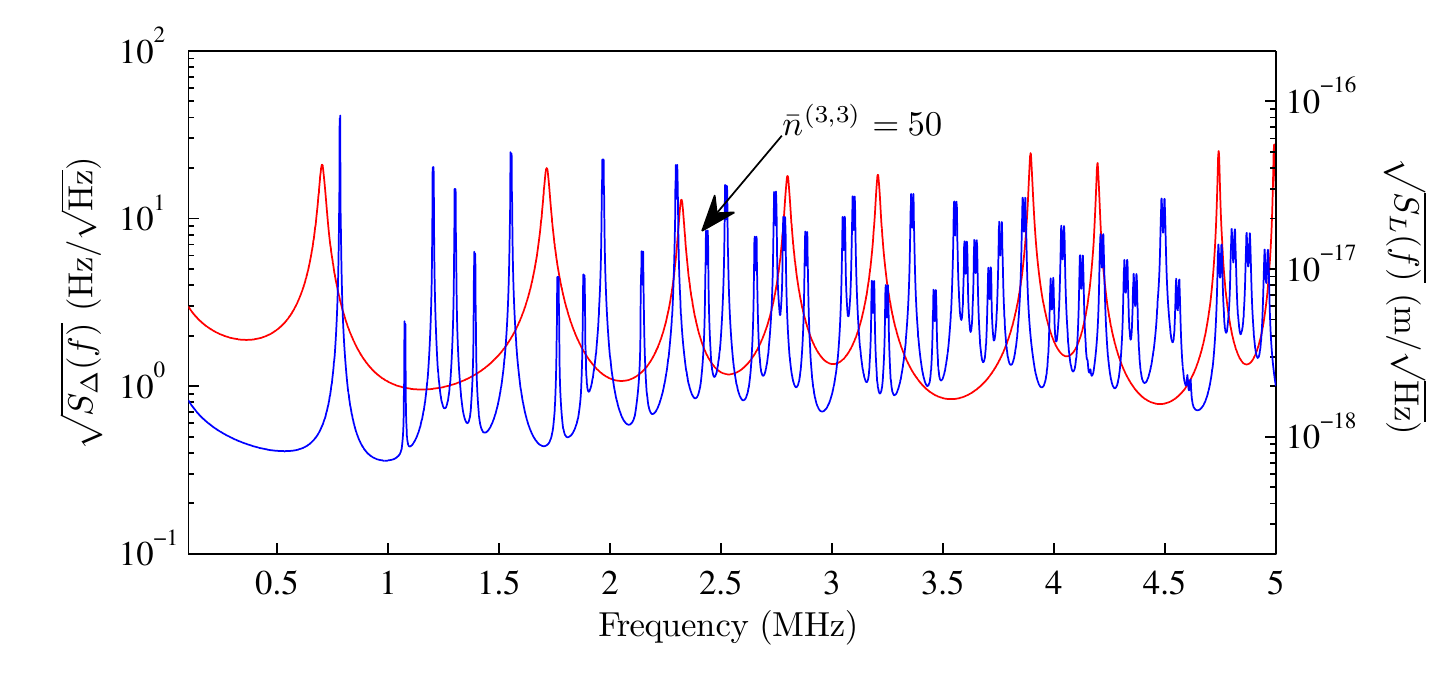}
\caption{Model spectrum of detuning fluctuations arising from mirror substrate (blue trace) and membrane motion (red trace) for the system described in Section \ref{se:Noise}.  The power and detuning of the cavity field are chosen so that the (3,3) membrane mode is optically damped to a thermal phonon occupation number of $\overline{n}^{(3,3)}= 50$.  The substrates vibrate at room temperature.}
\label{fig:Sub_Mem_Mod}
\end{figure}

In Figure \ref{fig:Sub_Mem_Mod} we show a numerical model of the thermal noise produced by an optically damped membrane  (Appendix \ref{ap:MemThermal}).  In the model we assume that each vibrational mode has a mechanical quality factor $Q_m=5\times 10^6$ and that the optical mode is centered on the membrane, so that only odd-ordered vibrational modes ($i = 1,3,5...; j = 1,3,5...$) are opto-mechanically coupled to the cavity (Appendix \ref{ap:Overlap}).  The power and detuning of the incident field have been chosen so that the  $(i,j)=(3,3)$ vibrational mode, with mechanical frequency $f_m^{(3,3)} = 2.32$ MHz,  is damped to a thermal phonon occupation number of  $\overline{n}^{(3,3)}= 50$.   Under these experimentally feasible conditions, we predict that the magnitude of $S_{\Delta}(f)$ produced by membrane thermal motion (blue curve) would be commensurate with the noise produced by substrate thermal motion (red curve).  Substrate thermal motion therefore constitutes an important a roadblock to observing quantum behavior in our system \cite{OmExp:Kimble2009}. 

\section{Strategy to suppress extraneous thermal noise}\label{se:Solution}

Extraneous thermal motion manifests itself as fluctuations in the cavity resonance frequency, and therefore the detuning of an incident laser field.  We now consider a method to suppress these detuning fluctuations using feedback.     Our strategy is to electro-optically imprint an independent measurement of the extraneous cavity resonance frequency fluctuations onto the frequency of the incident field, with gain set so that this added ``anti-noise'' cancels the thermal fluctuations.  To measure the extraneous noise, we monitor the resonance frequency of an auxiliary cavity mode which has nearly equal sensitivity to extraneous thermal motion but reduced (ideally no) sensitivity to thermal motion of the intracavity oscillator (further information could be obtained by  simultaneously monitoring multiple cavity modes).   The basis for this \emph{``differential sensitivity''} is the spatial overlap between the cavity modes and the vibrational modes of the optomechanical system (Appendix \ref{ap:Overlap}).  We hereafter specialize our treatment to the experimental system described in Section \ref{se:Noise}, in which case extraneous thermal motion corresponds to mirror ``substrate motion'' and motion of the intracavity oscillator to ``membrane motion'', respectively.  

 A conceptual diagram of the feedback scheme is shown in Figure \ref{fig:Sch}.  The field used for measurement of extraneous thermal noise is referred to as the ``probe field''.   The incident  field to which feedback is applied, and which is to serve the primary functions of the experiment, is referred to as the ``science field''.  The frequencies of the probe and science fields are $\nu_{0}^{p,s}=\mean{\nu_{0}^{p,s}}+\delta\nu_{0}^{p,s}$, respectively.  Each field is coupled to a single spatial mode of the cavity, referred to as the ``probe mode'' and the ``science mode'', respectively.  Resonance frequencies of the probe mode, $\nu_{c}^{p}$, and science mode, $\nu^{s}_{c}$, both fluctuate in time as a consequence of substrate motion and membrane motion.  We can represent these fluctuations, $\delta\nu_{c}^{p,s}\equiv \nu_{c}^{p,s} - \mean{\nu_{c}^{p,s}}$, as (Appendix \ref{ap:gandx})
\begin{equation}\label{eq:SubGeneral}
\begin{aligned}
\delta\nu_c^p &= \ g_1\delta x_1^p+g_2\delta x_2^p+g_{m}\delta x_{m}^p,\\
\delta\nu_c^s &= g_1\delta x_1^s+g_2\delta x_2^s+g_{m}\delta x_{m}^s.
\end{aligned}
\end{equation}
Here $\delta x_{1,2,m}^{p,s}$ denote the ``effective displacement'' of mirror substrate M1 (``1''), mirror substrate M2 (``2''), and the membrane (``m'') with respect to the probe (``p'') and science (``s'') cavity modes,  and $g_{1,2,m}$ denote the ``optomechanical coupling'' of M1, M2, and the membrane, respectively.  Effective displacement refers to the axial (along the cavity axis) displacement of the mirror or membrane surface averaged over the transverse intensity profile of the cavity mode (Eq. \eqref{eq:apparentdisplacement}).  Optomechanical coupling refers to the frequency shift per unit axial displacement if the entire surface were translated rigidly (Eq \eqref{eq:g}).  In the simple case for which the membrane is removed $(g_{m}=0)$,  couplings $g_{1,2}$ take on the familiar values: $g_1=-g_2=\mean{\nu_c^p}/\mean{L}\simeq \mean{\nu_c^s}/\mean{L}$.  Otherwise, all three are functions of the membrane's axial position relative to the intracavity standing-wave (Eqs. \eqref{eq:gmembrane}-\eqref{eq:gendmirrors}).  

To simplify the discussion of differential sensitivity, we confine our attention to a single vibrational mode of the membrane, with generalized amplitude $b_{m}$ and undamped mechanical frequency $f_{m}$  (Appendix \ref{ap:Overlap}).  We assume that cavity resonance frequencies $\nu_c^{p}$ and $\nu_c^{s}$ have different sensitivities to $b_{m}$ but are equally sensitive to substrate motion at Fourier frequencies near $f_{m}$.  We can express these two conditions in terms of the Fourier transforms \cite{PSD} of the effective displacements:
\begin{equation}\begin{aligned}
\delta x_{m}^{p,s}(f)&\equiv \eta_{p,s}b_{m}(f);\;\;\eta_{p}\ne \eta_{s}\\
\delta x_{1,2}^p(f)&\simeq \delta x_{1,2}^s(f)  \equiv \delta x_{1,2}(f).
\end{aligned}\label{eq:Assumption_x12}\end{equation}
Hereafter $ \eta_{p,s}$ will be referred to as ``spatial overlap'' factors.

\begin{figure}[t!]
\centering
\includegraphics[width=9cm]{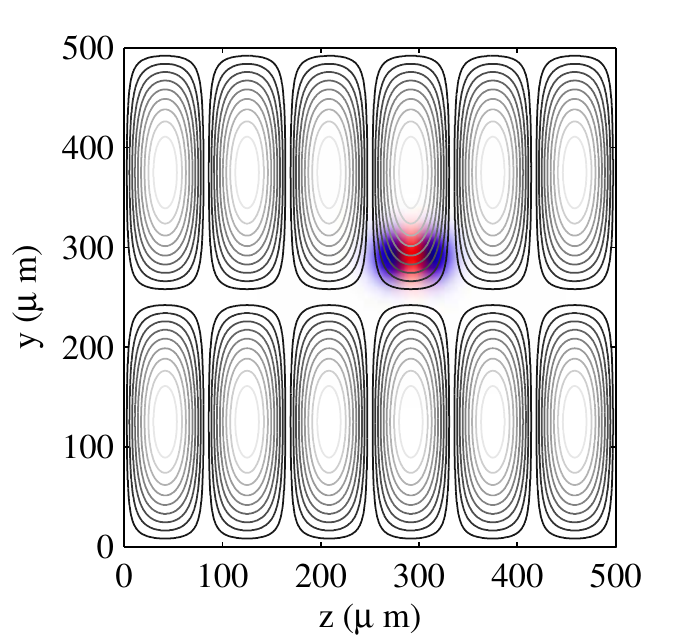}
\caption{ Location of the cavity modes relative to the membrane surface for experiments reported in Section \ref{subse:MIM_Exp} -- \ref{subse:2beam_exp}.  Density plots of the intra-cavity intensities of TEM$_{00}$ (red) and TEM$_{01}$ (blue) modes are displayed on top of a black contour plot representing the axial displacement of the (2,6) membrane mode.  Averaging the displacement of the surface weighted by the intensity profile gives the ``effective displacement'',$\delta x_m$, for membrane motion; in this case the effective displacement of the (2,6) mode is greater for the TEM$_{00}$ mode than it is for TEM$_{01}$ mode.}
\label{fig:BeamLocation}
\end{figure}

The first assumption of Eq. \eqref{eq:Assumption_x12} is valid if the vibrational mode shape of the membrane varies rapidly on a spatial scale set by the cavity waist size, $w_{0}$.  The latter assumption is valid if the opposite is true, i.e., we confine our attention to low order substrate vibrational modes, whose shape varies slowly on a scale set by $w_{0}$.  The substrate noise shown in Figure \ref{fig:Sub_Mod} fits this description, provided that the cavity mode is also of low order, e.g., cavity modes TEM$_{00}$ and TEM$_{01}$ (Eq. \eqref{eq:TEM}).  To visualize the differential sensitivity of TEM$_{00}$ and TEM$_{01}$, in Figure \ref{fig:BeamLocation} we plot the transverse intensity profile of each mode atop contours representing the amplitude of the (2,6) drum vibration of the membrane (Eq. \eqref{eq:membranemode}), with waist size and position and the membrane dimensions representing the experimental conditions discussed in Section \ref{subse:MIM_Exp}.  Choosing TEM$_{01}$ for the probe mode and TEM$_{00}$ for the science mode gives $\eta^{(2,6)}_{p}/\eta^{(2,6)}_{s}\approx 0.6$ for this example.

To implement feedback, a measurement of the probe field detuning fluctuations $\delta\Delta_p\equiv\delta\nu_0^p-\delta\nu_c^p$ is electro-optically mapped onto the frequency of the science field with gain $G$.
Combining Eqs. \eqref{eq:SubGeneral} and \eqref{eq:Assumption_x12}, and assuming that the laser source has negligible phase noise (i.e., $\delta\nu_0^p (f) =0$) and that $\delta\nu_0^s(f) = G(f)\delta\Delta_p(f)$, we can express the fluctuations in the detuning of the science field, $\delta\Delta_{s}\equiv\delta\nu_{0}^s-\delta\nu_c^{s}$, as
\begin{equation}
\begin{aligned}
\delta \Delta_{s}(f) &= G(f)\delta\Delta_p(f)-\delta\nu_c^{s}(f)\\
&=-\left(g_{1} \delta x_{1}(f)+g_{2} \delta x_{2}(f)\right)(1+G(f))-g_{m}\eta_{s}b_{m}(f)\left(1+(\eta_{p}/\eta_{s})G(f)\right).
\end{aligned}
\label{eq:feedback}\end{equation}
Here we have ignored the effect of feedback on the physical amplitude, $b_{m}$ (we consider this effect in Section \ref{se:Cooling}).

The science field detuning in Eq. \eqref{eq:feedback} is characterized by two components, an extraneous component proportional to $\left(1+G(f)\right)$ and an intrinsic component proportional to $\left(1+(\eta_{p}/\eta_{s})G(f)\right)$.   To suppress extraneous fluctuations, we can set the open loop gain to $G(f)=-1$.  The selectivity of this suppression is set by the ``differential sensing factor'' $\eta_{p}/\eta_{s}$.  In the ideal case for which the probe measurement only contains information about the extraneous noise, i.e. $\eta_{p}/\eta_{s}=0$, Eq. \eqref{eq:feedback} predicts that only extraneous noise is suppressed.  

For our open loop architecture, noise suppression depends critically on the phase delay of the feedback.  To emphasize this fact, we can express the open loop gain as 
\begin{equation}
G(f) = |G(f)|e^{2\pi if\tau(f)},
\label{eq:feedbackmagphase}\end{equation}
where $|G(f)|$ is the magnitude and $\tau(f)\equiv\text{Arg}[G(f)]/(2\pi f)$ is the phase delay of the open loop gain at Fourier frequency $f$.   Phase delay arises from the cavity lifetime and latencies in detection and feedback, and becomes important at Fourier frequencies for which $\tau(f)\gtrsim1/f$.   Since in practice we are only interested in fluctuations near the mechanical frequency of a single membrane mode, $f_{m}$, it is sufficient to achieve $G(f_{m}) =-1$ by manually setting $|G(f_{m})|=1$ (using an amplifier) and $\tau(f_{m})=j/(2f_m)$ (using a delay cable), where $j$ is an odd integer.  

Two additional issues conspire to limit noise suppression.   First, because substrate thermal motion is only partially coherent,  noise suppression requires that $\tau(f_{m})\ll Q/(2\pi f_{m})$, where $Q\sim 700$ is the quality of the noise peaks shown in figure \ref{fig:Sub_Mod}.   We achieve this by setting $\tau(f_{m})\sim1/(2f_{m})$.  Another issue is that any noise process not entering the measurement of $\delta\Delta_p$ via Eq. \eqref{eq:SubGeneral} will be \textit{added} onto the detuning of the science field via Eq. \eqref{eq:feedback}. The remaining extraneous contribution to $\delta\Delta_{s}$ will thus be nonzero even if $G(f)=-1$.   Expressing this measurement noise as an effective resonance frequency fluctuation $\delta\nu_{c}^N$, we can model the power spectrum of detuning fluctuations in the vicinity of $f_{m}$ as

\begin{equation}\begin{aligned}
S_{\Delta_{s}}(f) =&|1+G(f)|^2\cdot(g_{1}{}^2 S_{x_{1}}(f)+g_{2}{}^2 S_{x_{2}}(f))\\&+|1+(\eta_{p}/\eta_{s})^2G(f)|^2 \cdot g_{m}^2\eta_{p}^2 S_{b_{m}}(f)+|G(f)|^2\cdot S_{\nu_{c}^N}(f).
\end{aligned}\label{eq:feedbackPSD}\end{equation}

\begin{figure}[t!]
\centering
\subfigure[Zoomed out spectrum.]{
\includegraphics[width=11cm]{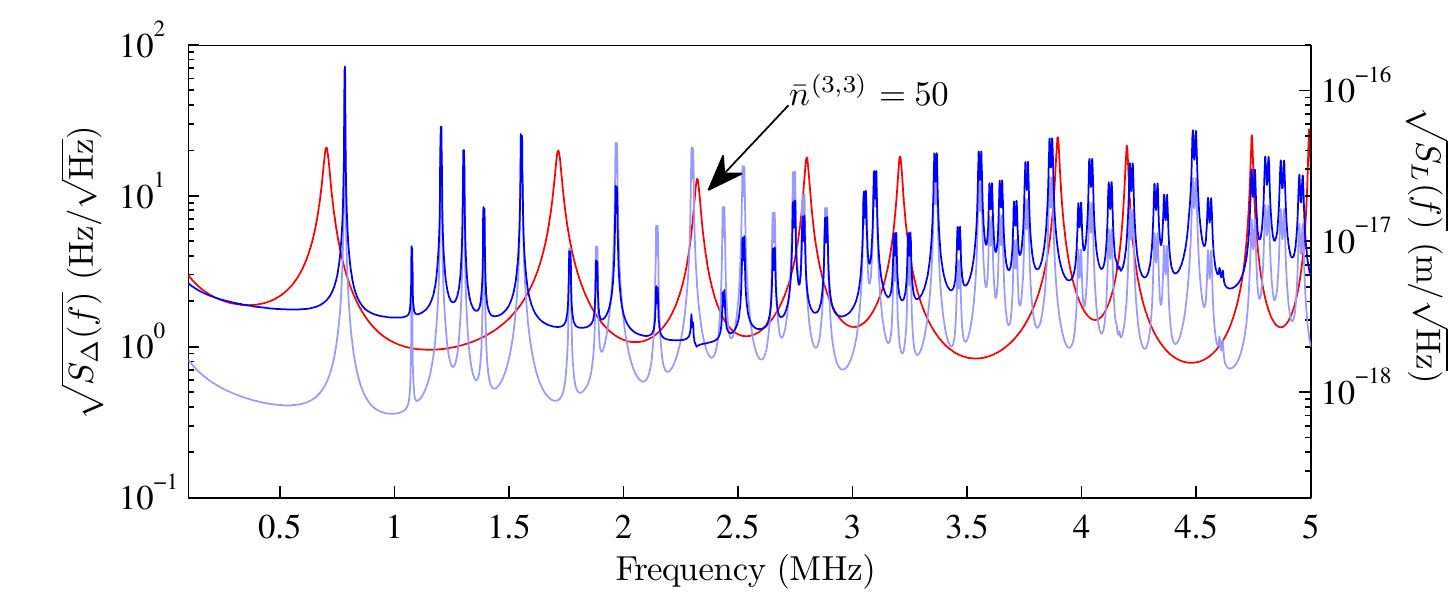}
\label{subfig:Supp_Model_ZoomOut}}
\subfigure[Zoomed in spectrum focusing on (3,3) membrane mode.]{
\includegraphics[width=11cm]{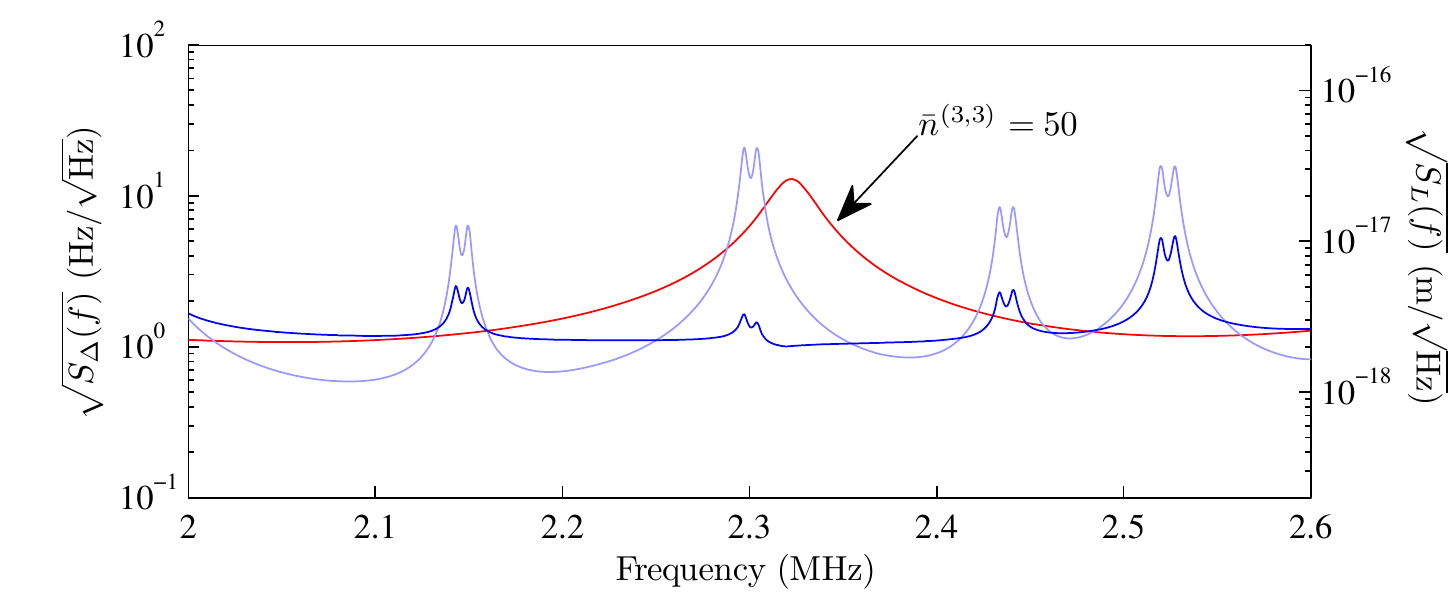}
\label{subfig:Supp_Model_ZoomIn}}
%\subfigure[Result:Membrane motion extracted]{
%\includegraphics[width=8cm]{Plots/2Probe_substrate_remove}
%}
\caption{Predicted suppression of substrate detuning noise (dark blue) for the science field based on a feedback with ideal differential sensing, $\eta_{p}/\eta_{s}=0$, for all modes, gain $G(f)=e^{\pi if/f_m^{(3,3)}}$, and measurement noise $S_{\nu_p^N} (f)= 1\;\text{Hz}^{2}/\text{Hz}$, where $f_m^{(3,3)}=2.32$ MHz is the mechanical frequency of the (3,3) membrane mode.  Unsuppressed substrate (light blue) and membrane noise (red) for the science field is taken from the model in Figure \ref{fig:Sub_Mem_Mod}.}
\label{fig:Sup_Sub_Mem_Mod}
\end{figure}

In Figure \ref{fig:Sup_Sub_Mem_Mod} we present an idealized model of our noise suppression strategy applied to the system described in Section \ref{se:Noise}.  We assume an ideal differential sensing factor of $\eta_{p}/\eta_{s}=0$ for all modes, a uniform gain magnitude of $|G(f)| = 1$, and a uniform phase delay $\tau(f) = 1/(2f_m^{(3,3)})$, where $f_m^{(3,3)}= 2.32 \text{ MHz}$ is the oscillation frequency of the (3,3) membrane mode.  As in Figure \ref{fig:Sub_Mem_Mod}, the detuning and power of the science field are chosen in order to optically damp the $(3,3)$ membrane mode to a thermal phonon occupation of $\overline{n}^{(3,3)}=50$. The probe measurement is also assumed to include extra noise at the level of $S_{\nu_c^N} (f)= 1\;\text{Hz}^{2}/\text{Hz}$.   Incorporating these assumptions into Eqs. \eqref{eq:feedbackmagphase} and \eqref{eq:feedbackPSD} produces the science field detuning spectrum shown in Figure \ref{fig:Sup_Sub_Mem_Mod}.  In this idealized scenario, substrate noise near the $(3,3)$ membrane mode is reduced to a level more than an order of magnitude below the peak amplitude of the (3,3) membrane mode.

\section{Experiment}\label{se:Exp}

\begin{figure}[t]
\centering
\includegraphics[width=11cm]{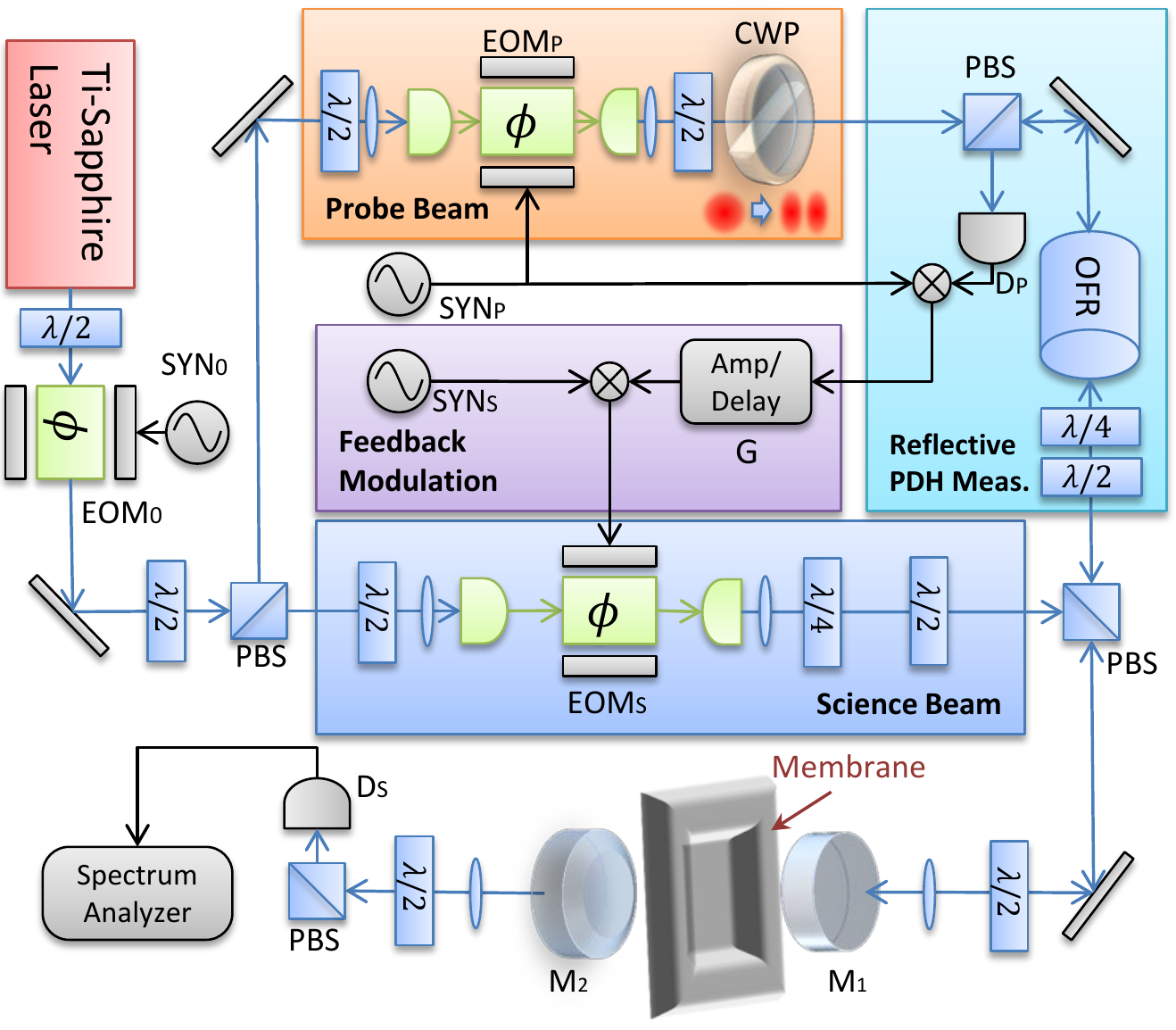}
\caption{Experimental setup: $\lambda/2$: half wave plate.  $\lambda/4$: quarter wave plate. PBS: polarizing beam splitter. EOM$_{0,P,S}$: electro-optical modulators for calibration and probe/science beams. CWP: split-$\pi$ wave plate. OFR: optical Faraday rotator. M$_{1,2}$: cavity entry/exit mirrors. D$_{P,S}$: photodetectors for probe/science beams. SYN$_{0,P,S}$: synthesizers for driving EOM$_{0,P,S}$.}
\label{fig:Sch_2beam}
\end{figure}

We have experimentally implemented the noise suppression scheme proposed in Section \ref{se:Solution}.  Core elements of the optical and electronic set-up are illustrated in Figure \ref{fig:Sch_2beam}.  As indicated, the probe and the science fields are both derived from a common Titanium-Sapphire (Ti-Sapph) laser, which operates at a wavelength of $\lambda_{0} \approx 810$ nm.  The science field is coupled to the TEM$_{00}$ cavity mode.  The probe field is coupled to either the TEM$_{00}$ or the TEM$_{01}$ mode of the cavity.  The frequencies of the science and probe fields are controlled by a pair of broadband electro-optic modulators (EOM$_{P,S}$ in Fig. \ref{fig:Sch_2beam}).  
% in combination with a single AOM (AOM$_{S}$) in the science beam path.

% Mode-matching to the TEM$_{01}$ mode is facilitated by passing the incident field through a split $\pi$ waveplate (a circular glass plate with a semi-circular $\pi$-phase coating). 

To monitor the probe-field detuning fluctuations $\delta\Delta_p$, the reflected probe field is directed to photodetector ``$D_{P}$'' and analyzed using the Pound-Drever-Hall (PDH) technique \cite{PDH}.
 The low frequency ($<$1 kHz) portion of the PDH signal is used to stabilize slow drift of $\nu_{p}$ via piezo-electric feedback to one of the end-mirrors.  The high frequency portion of the PDH signal is used to generate the electro-optic  feedback signal (Eq. \eqref{eq:feedback}).  Feedback gain $G(f)$ (Eq. \eqref{eq:feedbackmagphase}) is controlled by passing the PDH signal through a control box (``Amp/Delay''  in Figure \ref{fig:Sch_2beam}) containing an amplifier and a delay line.  The output of the control box, voltage $V_{\con}$, is used modulate the frequency of the science field using one of two methods.  The first method involves coupling $V_{\con}$ to the frequency-modulation (FM) port of the synthesizer (SYN$_{S}$) driving EOM$_{S}$.  The second method, not shown Fig. \ref{fig:Sch_2beam}, involves passing the science beam through an AOM driven by a voltage-controlled-oscillator (VCO), which is modulated by $V_{\con}$.  Feedback modifies the instantaneous detuning of the science field, $\Delta_{s}$, which we infer from the intensity of the transmitted field on photodetector ``$D_{S}$''.

We now develop several key aspects of the noise suppression scheme.   In Section \ref{subse:ExpBareCavity}, we emphasize the performance of the feedback network by suppressing substrate noise with the membrane removed from the cavity.  In Section \ref{subse:MIM_Exp}, we introduce the membrane and study the combined noise produced by membrane and substrate motion.  In Section \ref{subse:3beam_Exp}, we demonstrate the concept of differential sensing by electronically subtracting dual measurements of the probe and science mode resonance frequencies.   In Section \ref{subse:2beam_exp}, we combine these results to realize substrate noise suppression in the presence of the membrane.  We use a detuned science field for this study, and record a significant effect on the radiation pressure damping experienced by the membrane.  This effect is explored in detail in Section \ref{se:Cooling}.
% This effect is studied phenomologically in \ref{subse:PhenomenologicalCooling} and analytically in Section \ref{se:Cooling}

\subsection{Substrate noise suppression with the membrane removed}\label{subse:ExpBareCavity}

\begin{figure}[t!]
\centering
\includegraphics[width=13cm]{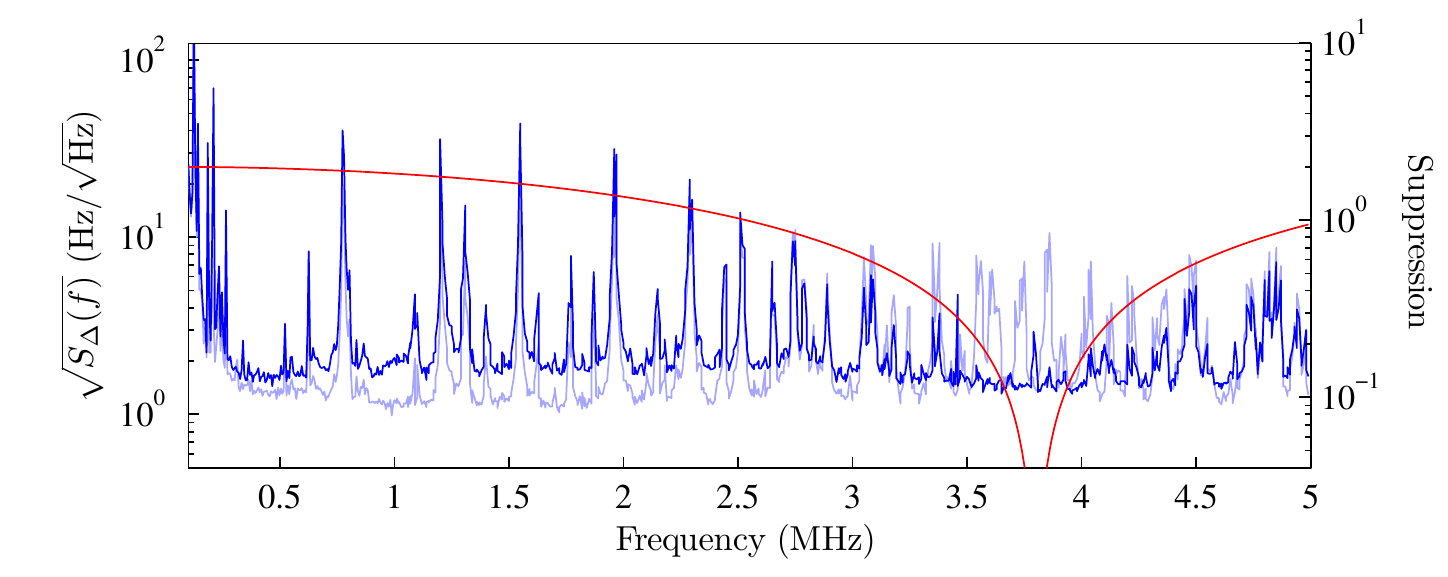}
\caption{Substrate noise suppression implemented with the membrane removed.  Gain is manually set to $G(f_{0}=3.8\; \mathrm{MHz})\approx -1$  using an RF amplifier and a delay line.  The ratio of the noise spectrum with (dark blue trace) and without (light blue trace) feedback is compared to the ``suppression factor'' $|1+G(f)|^2$ (red trace, right axis) with $G(f) = e^{\pi i f/f_{0}}$.}
\label{fig:Exp_Bare_Cavity}
\end{figure}

The performance of the feedback network is studied by first removing the membrane from the cavity, corresponding to $g_{m} = 0$ in Eqs. \eqref{eq:SubGeneral}, \eqref{eq:feedback}, and \eqref{eq:feedbackPSD}. The feedback objective is to suppress the detuning noise on a science field coupled to the TEM$_{00}$ cavity mode, shown for example in Figure \ref{fig:Sub_Mod}.  Absent the membrane, it is not necessary to employ a different probe mode to monitor the substrate motion.  For this example, both the science and probe field are coupled to the TEM$_{00}$ cavity mode.  The science field is coupled to one of the (nearly linear) polarization eigen-modes of  TEM$_{00}$ at a mean detuning $\langle \Delta_{s} \rangle =- \kappa$, where $\kappa \approx 4$ MHz is the cavity amplitude decay rate at 810 nm.  The probe field is resonantly coupled to the remaining (orthogonal) polarization eigen-mode of TEM$_{00}$.  Detuning fluctuations $\delta\Delta_{s}$ are monitored via the transmitted intensity fluctuations on detector $D_{S}$.    Detuning fluctuations $\delta\Delta_p$ are monitored via the PDH technique on detector $D_{P}$. 

Feedback is implemented by directing the measurement of $\delta\Delta_p$ to a VCO-controlled AOM in the science beam path (not shown in Fig. \ref{fig:Sch_2beam}).  The feedback gain $G(f)$ is tuned so that $S_{\Delta_{s}}(f)$ (Eq. \ref{eq:feedbackPSD}) is minimized at $f=f_{0}\approx$ 3.8 MHz, corresponding in this case to $|G(f_{0})|\approx1$ and $\tau(f_{0})=1/(2f_{0})$.   The magnitude of $S_{\Delta_{s}}(f)$ over a broad domain with (dark blue) and without (light blue) feedback is shown in Figure \ref{fig:Exp_Bare_Cavity}.  The observed suppression of $S_{\Delta_{s}}(f)$ may be compared to the predicted value of $|1+G(f)|^2$ based on a uniform gain amplitude and phase delay: i.e. $|G(f)|\approx1$ and $\tau(f)=1/(2f_{0})$ (Eq. \eqref{eq:feedbackmagphase}).   In qualitative agreement with this model (red trace in Figure \ref{fig:Exp_Bare_Cavity}), noise suppression is observed over a 3 dB bandwidth of $\sim 500$ kHz.  Noise suppression at target frequency $f_{0}$ is limited by shot noise in the measurement of $\delta\Delta_{p}$, corresponding to $S_{\nu_{c}^N}(f)\approx 1$ Hz$^2$/Hz in Eq. \eqref{eq:feedbackPSD}; this value was used for the model in Figure \ref{fig:Sup_Sub_Mem_Mod}.

\begin{figure}[t!]
\centering
\includegraphics[width=12cm]{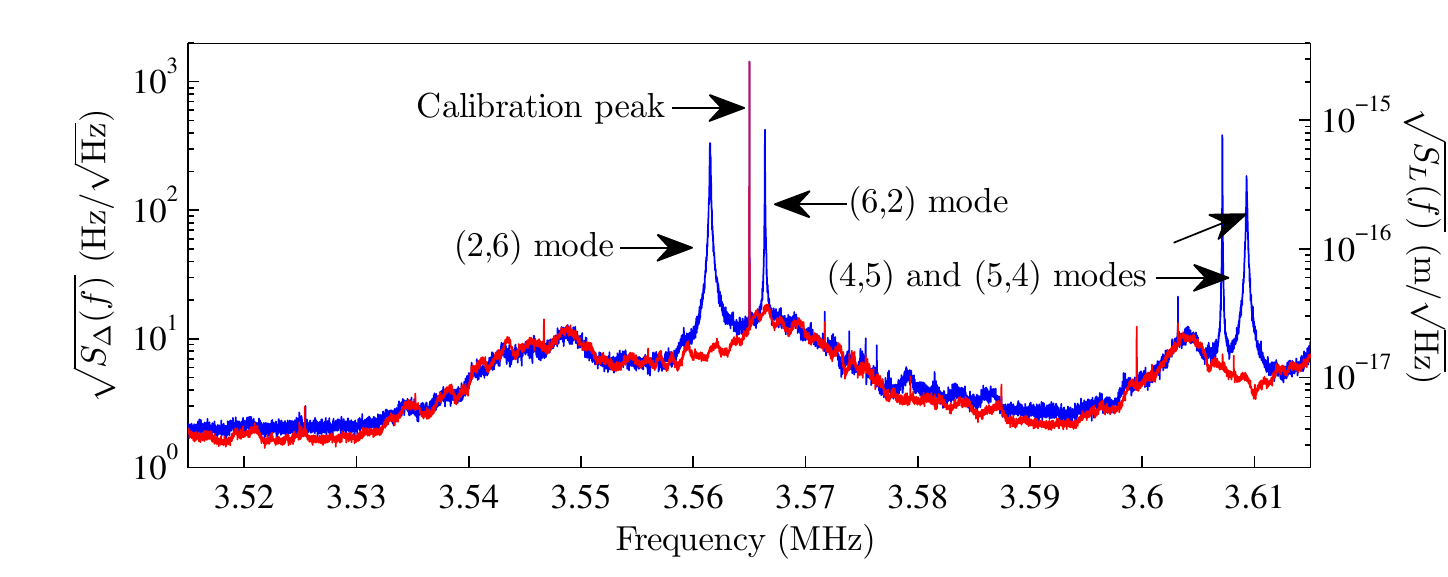}
\caption{Combined membrane and substrate thermal noise (blue trace) in the vicinity of $f_m^{(2,6)} = 3.56$ MHz, the frequency of the (2,6) vibrational mode of the membrane.  $g_{m}$ has been set to $\sim 0.04g_{1,2}$ in order to emphasize the substrate noise component.  For comparison, a measurement of the substrate noise with the membrane removed from the cavity is shown in red.  
%The calibration peak at 3.565 MHz is applied with EOM$_{0}$ in Figure \ref{fig:Sch_2beam}.
}
\label{fig:Sub_Mem_Exp}
\end{figure}

\subsection{Combined substrate and membrane thermal noise}\label{subse:MIM_Exp}

With the science field coupled to the TEM$_{00}$ cavity mode at $\langle \Delta_{s} \rangle = -\kappa$, we now introduce the membrane oscillator (described in Section \ref{se:Noise}).  We focus our attention on thermal noise in the vicinity of $f_m^{(2,6)}=3.56$ MHz, the undamped frequency of the (2,6) membrane vibrational mode.  To emphasize the dual contribution of membrane motion and substrate motion to fluctuations of $\nu_c^{s}$, we axially position the membrane so that $g_{m}$ has been set to $\sim 0.04 g_{1,2}$ (Appendix \ref{ap:gandx} and Eq. \eqref{eq:gm_Value}).  This reduces the fluctuations due to membrane motion, $g_{m}\delta x^{s}_{m}=g_{m}\eta_{s}^{(2,6)}b_{m}^{(2,6)}$ to near the level of the substrate noise $g_{1}\delta x_{1}+g_{2}\delta x_{2}$.  The location of the cavity mode relative to the displacement profile of the (2,6) mode has been separately determined, and is shown in Figure \ref{fig:BeamLocation}.  For $b_{m}$ coinciding with the amplitude of a vibrational antinode and the science mode coinciding with TEM$_{00}$, this location predicts a spatial overlap factor  (Eq. \eqref{eq:overlap}) of $\eta^{(2,6)}_{s}\approx 0.4$.

A measurement of the science field detuning noise, $S_{\Delta_{s}}(f)$,  with feedback turned off  ($G(f) = 0$) is shown in Figure \ref{fig:Sub_Mem_Exp}.  The blue trace shows the combined contribution of substrate and membrane thermal noise.  Note that the noise peaks associated with membrane motion are broadened and suppressed due to optical damping/cooling by the cavity field (Section \ref{se:Cooling}).  For comparison, we show an independent measurement made with the membrane removed (red trace).  Both traces were calibrated by adding a small phase modulation to the science field (EOM$_{0}$ in Fig. \ref{fig:Sch_2beam}).  We observe that the noise in the vicinity of $f_m^{(2,6)}$ contains contributions from multiple membrane modes and substrate modes.  The latter component contributes equally in both the blue and red traces, suggesting that substrate thermal motion indeed gives rise to the broad extraneous component in the blue trace.  From the red curve, we infer that the magnitude of the extraneous noise at $f_m^{(2,6)}$ is $S_{\Delta_{s},e}(f_m^{(2,6)}) \approx 80$ Hz$^2$/Hz (hereafter subscript ``e'' signifies ``extraneous'').  The influence of this background on the vibrational amplitude $b^{(2,6)}_{m}$ is discussed in Section \ref{se:Cooling}.

\subsection{Differential sensing of membrane and substrate motion}\label{subse:3beam_Exp}

\begin{table}[b]
\centering
\caption{Differential sensing factor, $\eta_{p}/\eta_{s}$, for the (2,6) and (6,2) membrane modes, with TEM$_{00}$ and TEM$_{01}$ forming the science and probe modes, respectively. The values in this table are inferred from Figure \ref{fig:3Beam_Result} and the model discussed Appendix \ref{ap:Overlap}.}\label{tab:DiffSensing}
\begin{tabular}{l| c |c}
\hline
Membrane mode & (2,6) & (6,2)\\
\hline
Determined from Figure \ref{fig:3Beam_Result} & 0.59 & 0.98\\
\hline
Calculated from Figure \ref{fig:BeamLocation} & 0.61 & 0.96\\
\hline 
\end{tabular}
\end{table}

To ``differentially sense'' the noise shown in Figure \ref{fig:Sub_Mem_Exp},  we use the probe field to monitor the resonance frequency of the TEM$_{01}$ mode.  Coupling the science field to the TEM$_{00}$ mode ($\nu_{s}$) and the probe field to the TEM$_{01}$ mode ($\nu_{p}$) requires displacing their frequencies by the transverse mode-splitting of the cavity, $f_{\tms}=\langle\nu_c^{p}\rangle-\langle\nu_c^{s}\rangle \approx$ 11 GHz ($f_{\tms}$ is set by the cavity length and the 5 cm radius of curvature of the mirrors).  This is done by modulating EOM$_{S}$ at frequency $f_{\tms}$, generating a sideband (constituting the science field) which is coupled to the TEM$_{00}$ mode when the probe field at the carrier frequency is coupled to the TEM$_{01}$ mode.   To spatially mode-match the incident Gaussian beam to the TEM$_{01}$ mode, the probe beam is passed through a split $\pi$ wave plate \cite{Split-Pi} (CWP in Figure \ref{fig:Sch_2beam}).  This enables a mode-matching efficiency of $\approx30\%$.

We can experimentally test the differential sensitivity  (Eq. \eqref{eq:Assumption_x12}) of modes TEM$_{00}$ and TEM$_{01}$  by electronically adding and subtracting simultaneous measurements of $\delta\Delta_p$ and $\delta\Delta_s$.  For this test, both measurements were performed using the PDH technique (detection hardware for the science beam is not shown in Figure \ref{fig:Sch_2beam}).  PDH signals were combined on an RF combiner after passing the science signal through an RF attenuator and a delay line.  The combined signal may be expressed as $\Delta_{\cmb}(f) =G_{0}(f)\delta\Delta_s(f)+\delta\Delta_p(f)$, where $G_{0}(f)$ represents the differential electronic gain.  

\begin{figure}[t!]
\centering
\subfigure[Zoomed out spectrum.]{
\includegraphics[width=12cm]{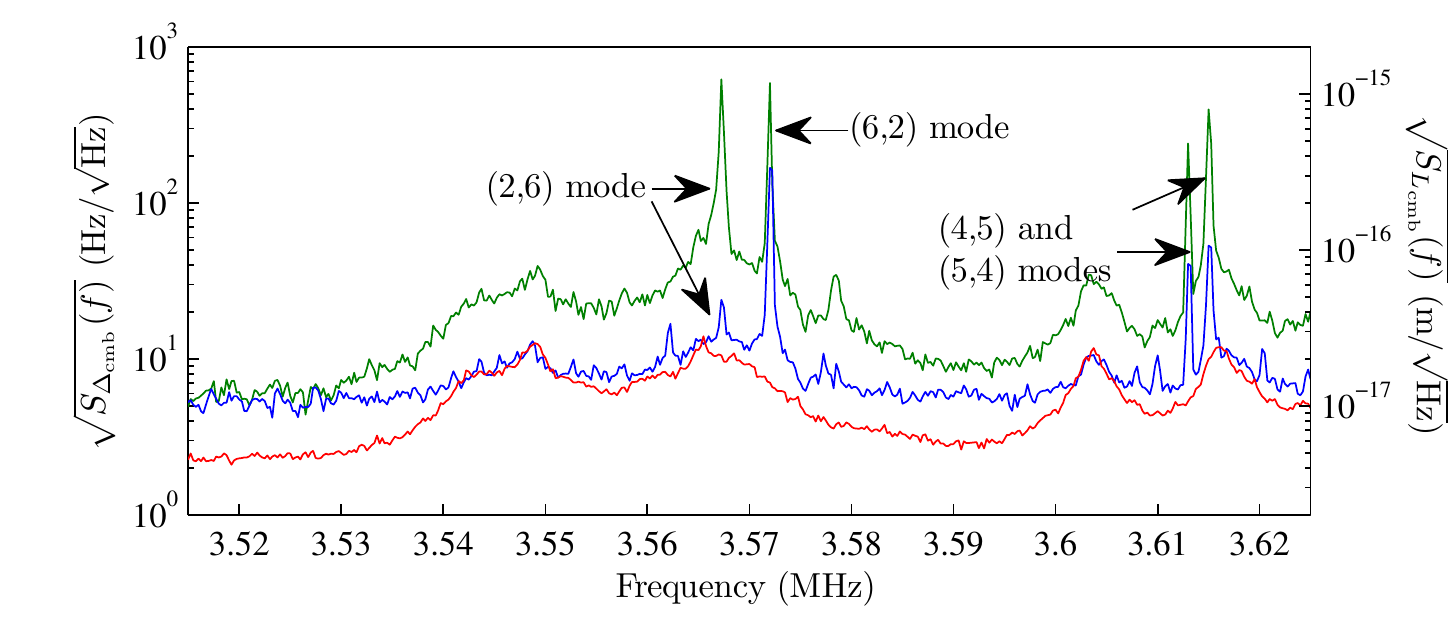}
\label{subfig:3Beam_ZoomOut}
}
\subfigure[Zoomed in spectrum.]{
\includegraphics[width=12cm]{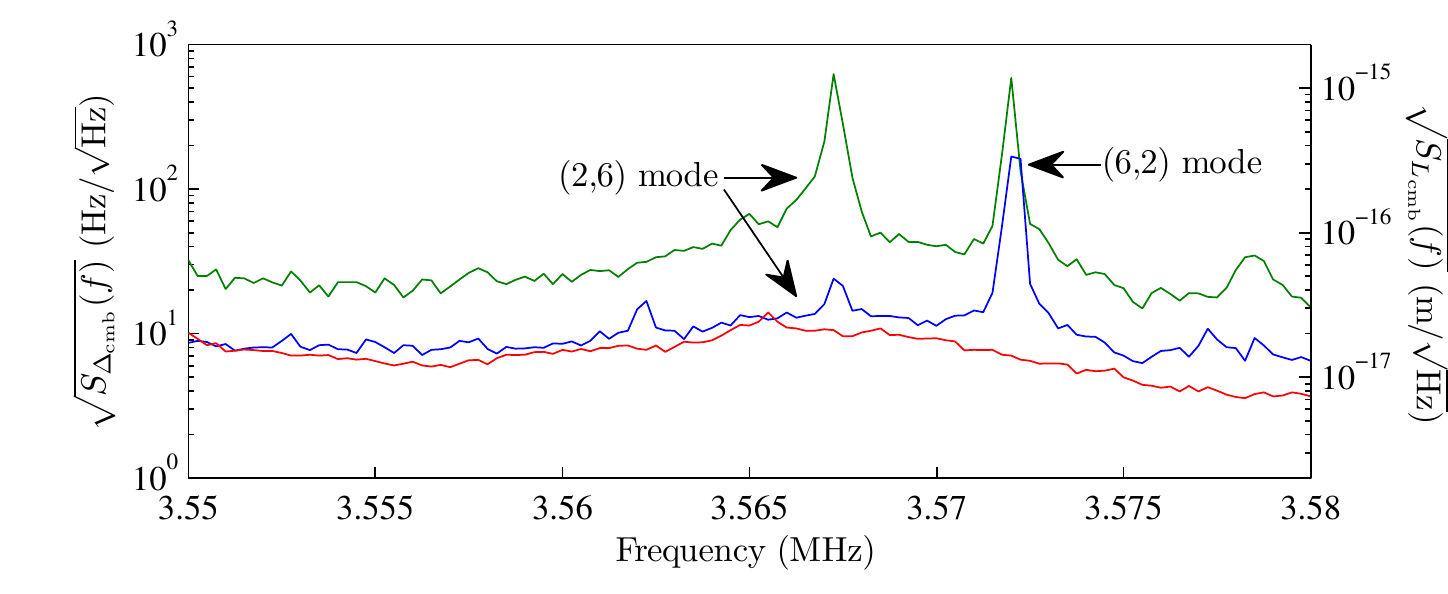}\label{subfig:3Beam_ZoomIn}}
\caption{Characterizing differential sensitivity of the TEM$_{00}$ (science) and TEM$_{01}$ (probe) mode to membrane motion.  Green and blue traces correspond to the noise spectrum of electronically added (green) and subtracted (blue) measurements of $\delta\Delta_p$ and $\delta\Delta_{s}$.  Electronic gain $G_{0}(f)$ has been set so that subtraction coherently cancels the contribution from the (2,6) mode at $f_m^{(2,6)}\approx3.568$ MHz.  The magnitude of the gain implies that $\eta_{p}^{(2,6)}/\eta_{s}^{(2,6)}\approx0.59$ and that $\eta_{p}^{(6,2)}/\eta_{s}^{(6,2)}\approx0.98$ for the nearby (6,2) noise peak at $f_{(6,2)} \approx 3.572$ MHz.} 
\label{fig:3Beam_Result}
\end{figure}

The power spectral density of the combined electronic signal \cite{PSD}, $S_{\Delta_{\cmb}}(f)$, is shown in Figure \ref{fig:3Beam_Result}, again focusing on Fourier frequencies near $f_m^{(2,6)}$.  In the blue (``subtraction'') trace, $G_{0}(f_m^{(2,6)})$ has been tuned in order to minimize the contribution from membrane motion, i.e. $G_{0}(f_m^{(2,6)})\approx-\eta^{(2,6)}_{p}/\eta^{(2,6)}_{s}$.  In the green (``addition'') trace, we invert this gain value.  Also shown (red) is a scaled measurement of the substrate noise made with the membrane removed, corresponding to the red trace in Figure \ref{fig:Sub_Mem_Exp} (note that the elevated noise floor in the green and blue traces is due to shot noise in the PDH measurements, which combine incoherently).   From the magnitude of the electronic gain, we can directly infer a differential sensing factor of  $\eta^{(2,6)}_{p}/\eta^{(2,6)}_{s}=0.59$ for the (2,6) membrane mode.  By contrast, it is evident from the ratio of peak values in the subtraction and addition traces that neighboring membrane modes (6,2), (4,5), and (5,4) each have different differential sensing factors.   For instance, the relative peak heights in Figure \ref{subfig:3Beam_ZoomIn} suggest that $\eta^{(6,2)}_{p}/\eta^{(6,2)}_{s} = 0.98$.  This difference relates to the strong correlation between $\eta_{p,s}$ and the location of the cavity mode on the membrane.  In Table \ref{tab:DiffSensing}, we compare the inferred differential sensing factor for (2,6) and (6,2) to the predicted value based on the cavity mode location shown in Figure \ref{fig:BeamLocation}.  These values agree to within a few percent.

\begin{figure}[t]
\centering
\subfigure[Zoomed out spectrum.]{
\includegraphics[width=11cm]{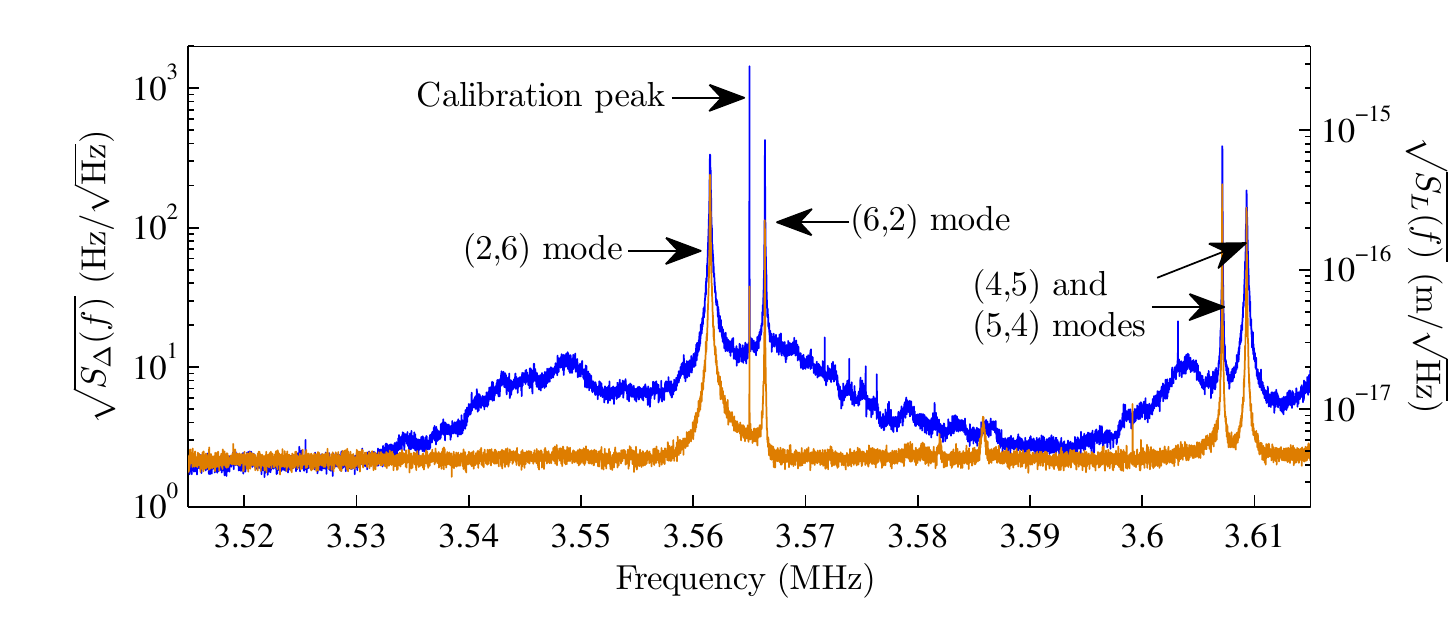}
\label{subfig:2Beam_ZoomOut}}
\subfigure[Zoomed in spectrum focusing on (2,6) membrane mode.]{
\includegraphics[width=11cm]{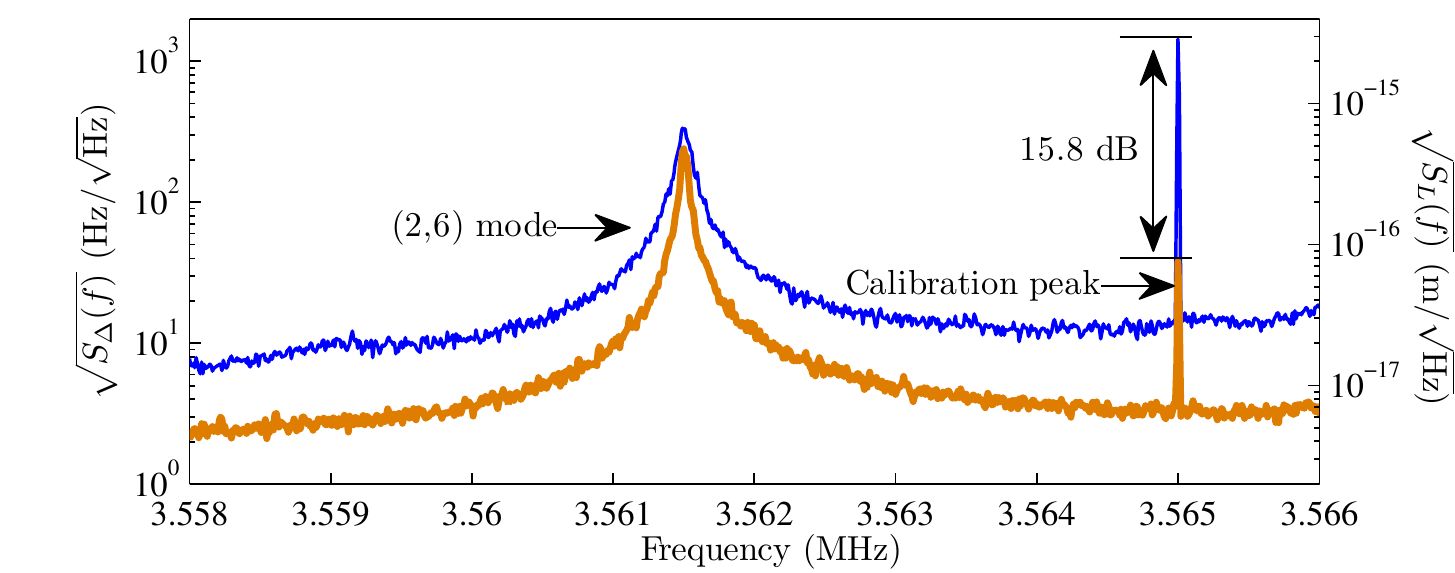}
\label{subfig:2Beam_ZoomIn}}
%\subfigure[Result:Membrane motion extracted]{
%\includegraphics[width=8cm]{Plots/2Probe_substrate_remove}
%}
\caption{Substrate noise suppression with the membrane inside the cavity.  Orange and blue traces correspond to the spectrum of science field detuning fluctuations with and without feedback, respectively. In order to suppress the substrate noise contribution near $f_m^{(2,6)}\approx3.56\text{ MHz}$, the feedback gain has been set to $G(f_m^{(2,6)})\approx-1$.  The feedback gain is fine-tuned by suppressing the detuning noise associated with an FM tone applied to both fields at 3.565 MHz.  The amplitude suppression achieved for this ``Calibration peak'' is 15.8 dB.}
\label{fig:2Beam_Result}
\end{figure}

\subsection{Substrate noise suppression with the membrane inside the cavity}\label{subse:2beam_exp}

Building upon Sections \ref{subse:ExpBareCavity}- \ref{subse:3beam_Exp}, we now implement substrate noise suppression with the membrane inside the cavity, which is the principal experimental result of our paper. The science field is coupled to the TEM$_{00}$ cavity mode with $\langle \Delta_{s}\rangle = -\kappa$, and $\delta\Delta_{s}$ is monitored via the transmitted intensity fluctuations on $D_{S}$. The probe field is coupled to the TEM$_{01}$ cavity mode, and $\delta\Delta_p$ is monitored via PDH on detector $D_{P}$.   Feedback is implemented by mapping the measurement of $\delta\Delta_p$ onto the frequency of the science field; this is done by modulating the frequency of the $\approx 11$ GHz sideband generated by EOM$_{S}$ (via the FM modulation port of synthesizer SYN$_{S}$ in Figure \ref{fig:Sch_2beam}).   The feedback objective is to selectively suppress the substrate noise component of $S_{\Delta_{s}}(f)$ near $f_m^{(2,6)}$ -- i.e., to subtract the red curve from the blue curve in Figure \ref{fig:Sub_Mem_Exp}.  To do this, the open loop gain of the feedback is set to $G(f_m^{(2,6)})\approx -1$ (Eq. \ref{eq:feedbackPSD}).  

The magnitude of $S_{\Delta_{s}}(f)$ with feedback on (orange) and off (blue) is shown in Figure \ref{fig:2Beam_Result}.   Comparing Figures \ref{fig:2Beam_Result} and \ref{fig:Sub_Mem_Exp}, we infer that feedback enables reduction of the substrate noise component at $f_m^{(2,6)}$ by a factor of $S_{\Delta_{s},e}(f_m^{(2,6)})|{}_{\text{ON}}/S_{\Delta_{s},e}(f_m^{(2,6)})|{}_{\text{OFF}} \approx 16$.   The actual suppression is limited by two factors: drift in the open loop gain and shot noise in the PDH measurement of $\delta\Delta_{p}$.  The first effect was studied by applying a common FM tone to both the probe and science field (via EOM$_{0}$ in Figure \ref{fig:Sch_2beam}; this modulation is also used to calibrate the measurements).  Suppression of the FM tone, seen as a noise spike at frequency $f_{0} = 3.565$ MHz in Fig. \ref{subfig:2Beam_ZoomIn}, indicates that the limit to noise suppression due to drift in $G(f)$ is $S_{\Delta_{s},e}(f_0)|{}_{\text{ON}}/S_{\Delta_{s},e}(f_{0})|{}_{\text{OFF}} \approx 1.4\times 10^3$.  The observed suppression of $\approx 16$ is thus limited by shot noise in the measurement of $\delta\Delta_p$.  This is confirmed by a small increase ($\sim$ 10\%) in $\sqrt{S_\Delta(f)}$ around $f=3.52$ MHz, and corresponds to $S_{\nu_{c}^{N}}(f)\sim1$ Hz$^2$/Hz in Eq. \eqref{eq:feedbackPSD}.  Note that the actual noise suppression factor is also partly obscured by shot noise in the measurement of $\delta\Delta_{s}$; this background is roughly $\sim 4$ Hz$^2$/Hz, coinciding with the level $S_{\Delta_{s}}(f)|{}_{\text{OFF}}$ (blue trace) at $f\approx 3.52$ MHz in Figure \ref{subfig:2Beam_ZoomOut}. 

In the following section, we consider the effect of electro-optic feedback on the membrane thermal noise component in Figure \ref{fig:2Beam_Result}.

\section{Extraneous noise suppression and optical damping: an application}\label{se:Cooling}

We now consider using a detuned science beam (as in Section \ref{subse:2beam_exp}) to optically damp the membrane.  Optical damping here takes place as a consequence of the natural interplay between physical amplitude fluctuations,  $b_{m}(f)$,  detuning fluctuations  $-g_{m}\eta_{s}b_{m}(f)$, and intracavity intensity fluctuations, which produce a radiation pressure force $\delta F_{\rad}(f) = -\varphi(f)\cdot g_{m}\eta_{s}b_{m}(f)$ that ``acts back'' on $b_{m}(f)$ (Appendix \ref{ap:suppress_length_frequency}).  The characteristic gain of this ``back-action'', $\varphi(f)$, possesses an imaginary component due to the finite response time of the cavity, resulting in mechanical damping of $b_{m}$ by an amount $\gamma_{\opt}\approx -\text{Im}[\delta F_{\rad}(f_{m})/b_{m}(f_{m})]/(2\pi f_{m} m_{\eff})$  (Eq. \eqref{eq:dampingandspring}), where $m_{\eff}$ is the effective mass of the vibrational mode (Eq. \eqref{eq:effmass}).

In our noise suppression scheme, electro-optic feedback replaces the intrinsic detuning fluctuations, $-g_{m}\eta_{s}b_{m}(f)$, with the modified detuning fluctuations, $-(1+(\eta_{p}/\eta_{s})G(f))g_{m}\eta_{p}b_{m}(f)$ (Eq. \eqref{eq:feedback}).  The radiation pressure force experienced by the membrane is thus modified by a factor of  $(1+(\eta_{p}/\eta_{s})G(f))$ (this reasoning is analytically substantiated in Appendix \ref{ap:suppress_length_frequency}). Here we define a parameter $\mu \equiv (\eta_{p}/\eta_{s})G(f_{m})$.  For purely real $G(f)$, the modified optical damping rate as a function of $\mu$ (where $\mu=0$ in the absence of electro-optic feedback) has the relation (Eq. \eqref{eq:dampingandspring}),
\begin{equation}
\frac{\gamma_{\opt}(\mu)}{\gamma_{\opt}(\mu=0)}\approx 1+\mu.
\label{eq:opticaldamping}
\end{equation}

Associated with optical damping is ``optical cooling'', corresponding to a reduction of the vibrational energy from its equilibrium thermal value.  From a detailed balance argument  it follows that \cite{OmRev:VahalaOE07}
\begin{equation}
\langle b_{m}^2 \rangle(\mu=0) = \frac{\gamma_{m}}{\gamma_{m}+\gamma_{\opt} (\mu=0)} \frac{k_{B}T_{b}}{m_{\eff}(2\pi f_{m})^2},
\label{eq:opticalcooling_NoFB}\end{equation}
where $k_{B}$ is the Boltzmann constant and $T_{b}$ is the temperature of the thermal bath. In Appendices \ref{ap:MemThermal} and \ref{ap:suppress_length_frequency}, we extend Eq. \eqref{eq:opticalcooling_NoFB} to the case with feedback and find the modified expression
\begin{equation}
\langle b_{m}^2 \rangle(\mu) = \frac{\gamma_{m}}{\gamma_{\eff}(\mu)} \frac{k_{B}T_{b}}{m_{\eff}(2\pi f_{m})^2},
\label{eq:opticalcooling}\end{equation}
where $\gamma_{\eff} (\mu) \equiv \gamma_{m}+\gamma_{\opt}(\mu)$ is the effective mechanical damping rate.  

From Eqs. \eqref{eq:opticaldamping} and \eqref{eq:opticalcooling}, we predict that when the probe is insensitive to membrane motion ($\eta_{p}=0\rightarrow\mu=0$), optical damping/cooling is unaffected by electro-optic feedback.  In a realistic scenario for which $(\eta_{p}/\eta_{s})>0$ (e.g., Section \ref{se:Exp}), feedback with $G(f_{m}) = -1$ (to suppress extraneous noise) results in a reduction of the optical damping.  Remarkably, when $(\eta_{p}/\eta_{s})<0$, extraneous noise suppression can coincide with increased optical damping (i.e., reduced phonon number in the presence of extraneous noise suppression), which  we elaborate Section \ref{subse:OC}.

It is worth emphasizing that the effect described in Eq. \eqref{eq:opticaldamping} has much in common with active radiation pressure feedback damping, a.k.a ``cold-damping'', as pioneered in \cite{OmFbExp:PinardCool99}.  In the experiment described in \cite{OmFbExp:PinardCool99}, feedback is applied to the position of a micro-mirror (in a Fabry-P\'{e}rot cavity) by modulating the intensity of an auxiliary laser beam reflected from the mirror's surface.  This beam imparts a fluctuating radiation pressure force, which may be purely damping (or purely anti-damping) if the delay of the feedback is set so that the intensity modulation is in phase (or $\pi$ out of phase) with the oscillator's velocity. 

Our scheme differs from \cite{OmFbExp:PinardCool99} in several important ways. In \cite{OmFbExp:PinardCool99}, the ``probe'' field is coupled to the cavity,  but the intensity modulated ``science'' field does not directly excite a cavity mode.  By contrast, in our scheme, both the probe and science fields are coupled to independent spatial modes of the cavity.  Moreover, instead of directly modifying the intensity of the incident science field, in our scheme we modify its detuning from the cavity, which indirectly modifies the \emph{intra-cavity} intensity.  The resulting radiation pressure fluctuations produce damping (or anti-damping) of the oscillator (in our case a membrane mode) if the detuning modulation is in phase (or $\pi$ out of phase) with the oscillator's velocity.  For the extraneous noise suppression result shown in Figure \ref{fig:2Beam_Result}, the phase of the electro-optic feedback results in anti-damping of the membrane's motion.  The reason why this is tolerable, and another crucial difference between our scheme and the ``cold-damping'' scheme of  \cite{OmFbExp:PinardCool99}, is that the feedback force is super-imposed onto a strong cavity ``back-action'' force.  In Figure \ref{fig:2Beam_Result}, for example, the small amount of feedback anti-damping is negated by larger, positive back-action damping.  The relative magnitude of these two effects depends on the differential sensing factor ($\eta_{p}/\eta_{s}$) for the two cavity modes.

To investigate the interplay of electro-optic feedback and optical damping, we now reanalyze the experiment described in Section \ref{subse:2beam_exp}.   In that experiment, the science field was red-detuned by $\langle \Delta_{s}\rangle \approx -\kappa\approx-4$ MHz, resulting in significant damping of the membrane motion.  This damping is evident in a careful analysis of the width $\gamma_{\eff}$ (FWHM in Hz$^2$/Hz units) and area $\langle \delta\Delta_{s}^{2} \rangle \equiv\int_{f_{m}}S_{\Delta_{s}}(f)\mathrm{d}f$ of the thermal noise peak centered near $f_{m}=f_m^{(2,6)}$ in Fig. \ref{fig:2Beam_Result}.  We have investigated the influence of electro-optic feedback on optical damping by varying the magnitude of the feedback gain, $G(f_m^{(2,6)})$, while monitoring $S_{\Delta_{s}}(f)$ in addition to $S_{\Delta_{p}}(f)$ (inferred from the probe PDH measurement). From the basic relations given in Eqs. \eqref{eq:opticaldamping} and \eqref{eq:opticalcooling}, the ratio of the effective damping rate with and without electro-optic feedback is predicted to scale linearly with $\mu$, i.e.,
\begin{equation}
R_{\gamma_\eff}(\mu)\equiv\frac{\gamma_{\eff}(\mu)}{\gamma_{\eff}(\mu=0)}= 1+\frac{\gamma_{\eff}(\mu=0)-\gamma_{m}}{\gamma_{\eff}(\mu=0)}\mu.
\label{eq:ratiogamma}\end{equation}

Similarly, combining Eqs. \eqref{eq:SubGeneral} and \eqref{eq:opticalcooling} and ignoring substrate noise, we predict fluctuations $\delta\Delta_{p}$ to have the property
\begin{equation}
R_{\Delta_p}(\mu)\equiv\frac{\langle \delta\Delta_{p}^2\rangle(\mu=0)}{\langle \delta\Delta_{p}^{2}\rangle(\mu)} = R_{\gamma_\eff}(\mu),
\label{eq:ratioprobe}\end{equation}
where $\langle \delta\Delta_{p}^{2}\rangle = \int_{f_{m}}S_{\Delta_{p}}(f)\mathrm{d}f$ is the area beneath the noise peak centered at $f_{m}$.

Finally, using Eq. \eqref{eq:feedbackPSD}, the fluctuating detuning of science field is predicted to obey 
\begin{equation}
R_{\Delta_s}(\mu)\equiv\frac{\langle \delta\Delta_{s}^2\rangle(\mu=0)}{\langle \delta\Delta_{s}^{2}\rangle(\mu)} = \frac{R_{\gamma_\eff}(\mu)}{(1-\mu)^2},
\label{eq:ratioscience}\end{equation}
where $\langle\delta\Delta_{s}^2\rangle = \int_{f_{m}}S_{\Delta_{s}}(f)\mathrm{d}f$.

\begin{figure}[t!]
\centering
\includegraphics[width=12cm]{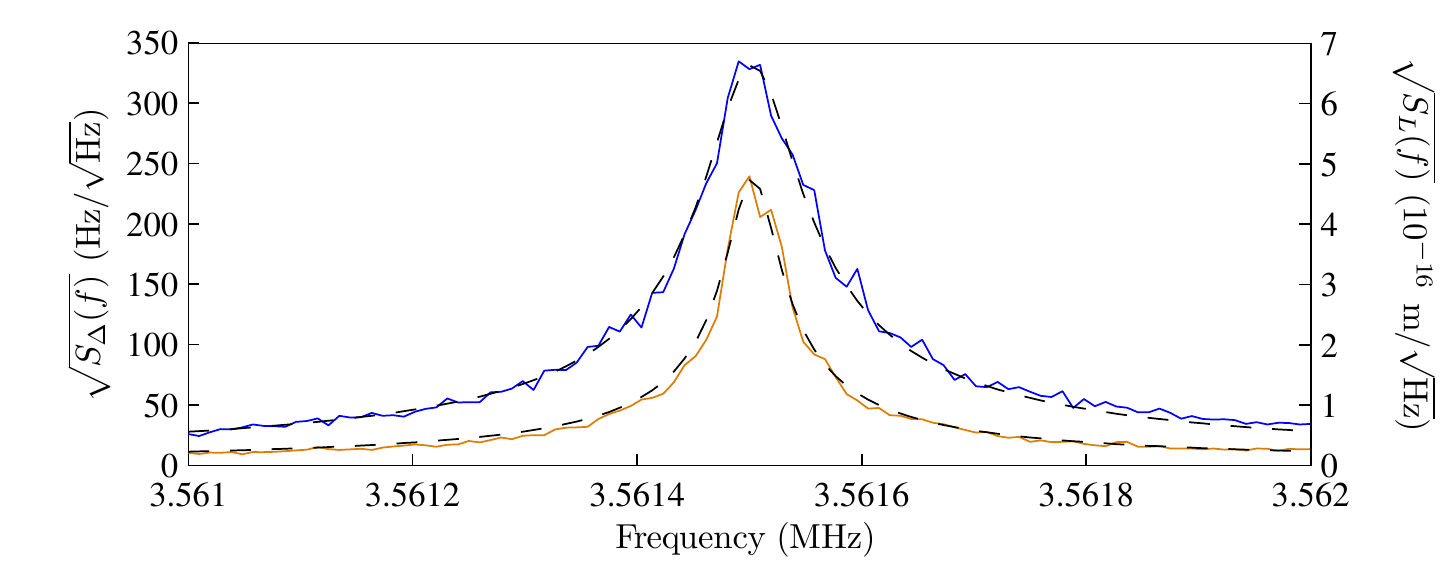}
\caption{Lorentzian fits of the thermal noise peak near $f_m^{(2,6)}$ in Figure \ref{fig:2Beam_Result}, here plotted on a linear scale.  Solid blue and orange traces correspond to science field detuning noise with noise suppression off and on, respectively. Dashed traces correspond to Lorentzian fits.}
\label{fig:FeedOnOffFit}
\end{figure}

\begin{table}[b!]
\center
\caption{Parameters from Figures \ref{fig:Sub_Mem_Exp} and \ref{fig:FeedOnOffFit}. $\mu=0$ and $\mu=-\eta_{p}^{(2,6)}/\eta_{s}^{(2,6)}$ represents the noise suppression is off and on, respectively.  $\gamma_\text{eff}$ and  $\sqrt{\mean{\delta\Delta_{s}^2}}$ are inferred from the Lorentzian fits in Figure \ref{fig:FeedOnOffFit}. $S_{\Delta_{s},e}(f_m^{(2,6)})$ with $\mu=0$ and $\mu=-\eta_{p}^{(2,6)}/\eta_{s}^{(2,6)}$ are inferred from the red curve in Figure \ref{fig:Sub_Mem_Exp} and the orange curve in Figure \ref{fig:2Beam_Result}, respectively.}\label{tab:2BeamPara}
\begin{tabular}{c| c| c|c}
\hline
$\mu$& $\gamma_\text{eff}$ (Hz) & $\sqrt{\mean{\delta\Delta_{s}^2}}$ (Hz)& $S_{\Delta_{s},e}(f_m^{(2,6)})$ (Hz$^2$/Hz)\\
\hline
0 & $ 21$ & $ 3.8\times 10^3$ & $80$\\
\hline
$-\eta_{p}^{(2,6)}/\eta_{s}^{(2,6)}$&$ 12$ & $2.1\times 10^3$ & $5$\\
\hline
\end{tabular}
\end{table}

To obtain values for $\gamma_{\eff}$, $\langle\delta\Delta_{p}^2\rangle$,  and  $\langle\delta\Delta_{s}^2\rangle$, we fit the noise peak near $f_m^{(2,6)}$ in measurements of $S_{\Delta_{p}}(f)$ and $S_{\Delta_{s}}(f)$ to a Lorentzian line profile (Appendix \ref{ap:MemThermal}).  Two examples, corresponding to the noise peaks in Figure \ref{subfig:2Beam_ZoomIn}, are highlighted in Figure \ref{fig:FeedOnOffFit}.  The blue curve corresponds to $S_{\Delta_{s}}(f)$ with $\mu = 0$ ($G(f_m^{(2,6)}) = 0$) and the orange curve with $\mu = -\eta_{p}^{(2,6)}/\eta_{s}^{(2,6)}$ ($G(f_m^{(2,6)}) = -1$).  Values for $\gamma_{\eff}$ and $\langle\delta\Delta_{s}^2\rangle$ inferred from these two fits are summarized in Table \ref{tab:2BeamPara}.   Using these values and a separate measurement of $\gamma_{m}^{(2,6)} = 4.5$ Hz, we can test the model by comparing the differential sensing factor inferred from  Eq. \eqref{eq:ratiogamma} and Eq. \eqref{eq:ratioscience}.  From Eq. \eqref{eq:ratiogamma} we infer $\eta_{p}^{(2,6)}/\eta_{s}^{(2,6)}= 0.54$ and from Eq. \eqref{eq:ratioscience} we infer $\eta_{p}^{(2,6)}/\eta_{s}^{(2,6)} \approx 0.61$.  These values agree to within $10\%$ of each other and the values listed in Table \ref{tab:DiffSensing}.

\begin{figure}[t]
\centering
\includegraphics[width=13cm]{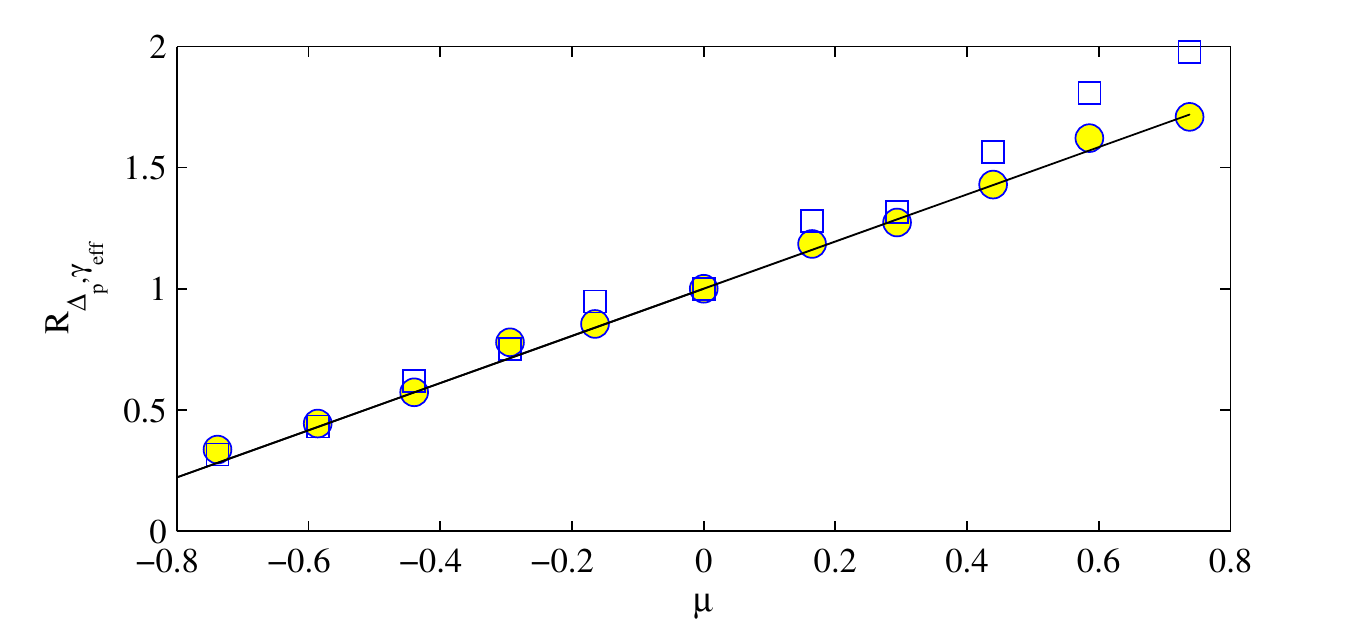}
\caption{The impact of electro-optic feedback on optical damping/cooling of the (2,6) membrane mode with a red-detuned science field,  as reflected in measured ratios $R_{\gamma_\eff}$ (yellow circles, Eq. \eqref{eq:ratiogamma}) and $R_{\Delta_p}$ (blue squares, Eq. \eqref{eq:ratioprobe}), as a function of feedback gain parameter $\mu$.  The model shown (black line) is for  $\eta_{s}^{(2,6)}/\eta_{p}^{(2,6)} = 0.6.$}
\label{fig:FeedbackRatio}
\end{figure}

In Figure \ref{fig:FeedbackRatio} we show measurements of $R_{\gamma_\eff}$ (yellow circles) and $R_{\Delta_p}$ (blue squares) for several values of $\mu$, varied by changing the magnitude of $G(f_m^{(2,6)})$.  The horizontal scale is calibrated by assuming $\eta_{s}^{(2,6)}/\eta_{p}^{(2,6)} = 0.6$.  Both measured ratios have an approximately linear dependence on $\mu$ with a common slope that agrees with the prediction based on Eqs. \eqref{eq:ratiogamma} and \eqref{eq:ratioprobe} (black line).

\section{Discussion}\label{se:Discussion}

\subsection{Optical cooling limits}

Taking into account the reduction of extraneous noise (Section \ref{subse:2beam_exp}) and the effect of electro-optic feedback on optical damping (Section \ref{se:Cooling}), we now estimate the base temperature achievable with optical cooling in our system.  Our estimate is based on the laser frequency noise heating model developed in \cite{OmTheory:RablPhaseNoise09}, in which it was shown that the minimum thermal phonon occupation achievable in the presence of laser frequency noise  $S_{\nu_{0}}(f)$ is given by
\begin{equation}
\overline{n}_{\min}\simeq \frac{2 \sqrt{\Gamma_mS_{\nu_{0}} (f_{m})}}{g \delta x_\text{zp}}+\frac{\kappa^2}{f_{m}^2},
\label{eq:basetemperature}\end{equation}
where $\Gamma_{m}=k_{B}T_{b}/h f_{m}$ is the re-thermalization rate at environment temperature $T_{b}$ and $g\delta x_\text{zp}$ is the cavity resonance frequency fluctuation associated with the oscillator's zero point motion.

\begin{figure}[t]
\centering
\includegraphics[width=13cm]{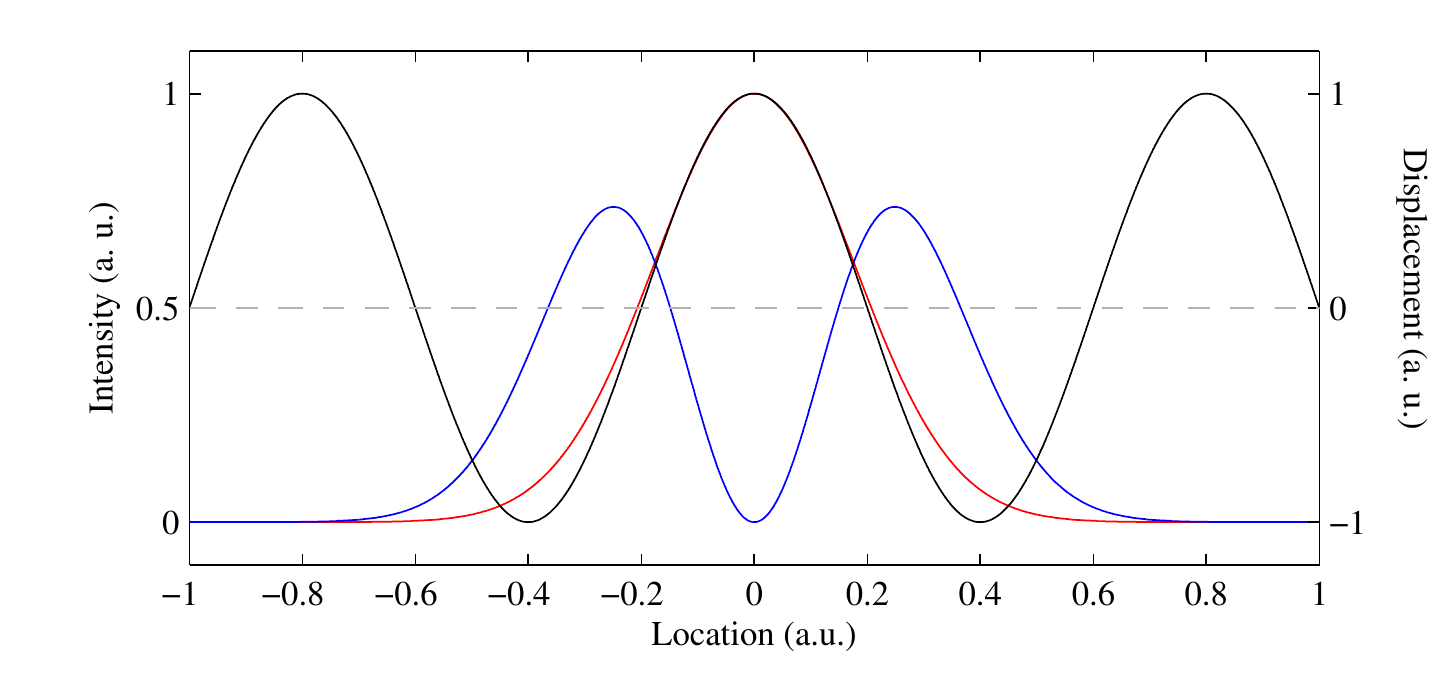}
\caption{Visualization of ``negative'' differential sensing.  The transverse displacement profile of the (1,5) membrane mode is shown in black (a 1D slice along the midline of the membrane is shown).  Red and blue curves represent the transverse intensity profile of TEM$_{00}$ and TEM$_{01}$ cavity modes, both centered on the membrane.  The cavity waist size is adjusted so that the displacement averaged over the intensity profile is negative for TEM$_{01}$ and positive for TEM$_{00}$.}
\label{fig:OC}
\end{figure}

We apply Eq. \eqref{eq:basetemperature} to our system by replacing $S_{\nu_{0}}(f_{m})$ with extraneous detuning noise, $S_{\Delta_{s},e}(f_{m})$, and $g\delta x_\text{zp}$ with feedback-modified zero-point fluctuations $g_{m}(1-\eta_{p}/\eta_{s})\eta_{s}b_\text{zp}$, where $b_\text{zp} = \sqrt{\hbar/(4\pi f_{m} m_{\eff})}$ and $m_{\eff}$ are the zero-point amplitude and effective mass of amplitude coordinate $b_{m}$, respectively (Appendix \ref{ap:MemThermal}).  We also assume that the membrane can be positioned so as to increase the opto-mechanical coupling to its maximal value $g_{m}^{\max} = 2|r_{m}|\langle \nu_c^{s} \rangle/\langle L \rangle$ ($r_{m}$ is the membrane reflectively, Appendix \ref{ap:gandx}) without effecting the magnitude of the suppressed substrate noise.  
%Estimated final occupation numbers with and without feedback are summarized in Table \ref{tab:2BeamPara}.  
Values for $S_{\Delta_{s},e}(f_m^{(2,6)})$ with feedback on and off are drawn from Table \ref{tab:2BeamPara}: $5$ Hz$^2$/Hz and $80$ Hz$^2$/Hz, respectively.  Other parameters used are $\kappa=4$ MHz, $f_{m}=f_m^{(2,6)}=3.56$ MHz, $m_{\eff}=m_{p}/4 = 8.4$ ng, $|r_{m}|=0.465$ (\cite{OmExp:HarrisDispersive08}),  $g_{m}^{\max} = 4.6\times10^5$ MHz/$\mu$m (for $\langle L \rangle$ = 0.74 mm), $\eta_{p}^{(2,6)} =0.24$, and $\eta_{s}^{(2,6)} = 0.4$ (from Fig. \ref{fig:BeamLocation}).  With and without feedback, we obtain values of $\overline{n}^{(2,6)}_{\min}$ = 128 and 206, respectively.  

The improvement from $\overline{n}^{(2,6)}_{\min}$ = 206 to 128 is modest and derives from the use of a relatively large and positive differential sensing factor, $\eta_{p}^{(2,6)}/\eta_{s}^{(2,6)}\approx 0.6$, as well as shot noise in the probe measurement, which sets a lower bound for the extraneous detuning fluctuations ($S_{\Delta_{s},e}(f_m^{(2,6)})\approx S_{\nu_{c}^{N}}(f_m^{(2,6)})\approx 5$ Hz$^2$/Hz in Figure \ref{fig:2Beam_Result}). 
 Paths to reduce or even change the sign of the differential sensing factor are discussed in Sections \ref{subse:OC}.  Reduction of shot noise requires increasing the technically accessible probe power; as indicated in Section \ref{subse:2beam_exp}, we suspect that significant reduction of $S_{\Delta_{s},e}(f_m^{(2,6)})$ can be had if this improvement were made.

Note that the above values of $\overline{n}^{(2,6)}_{\min}$ are based on an estimate of the maximum obtainable opto-mechanical coupling $g_{m}^{\max}$.  For the experiment in Section \ref{se:Exp}, however, we have set the opto-mechanical coupling coefficient $g_m\ll g_{m}^{\max}$ in order to emphasize the substrate noise.  From the data in Figure \ref{fig:2Beam_Result}, we infer $g_m$ to be
\begin{equation}\label{eq:gm_Value}
g_m= \frac{2\pi f_m^{(2,6)}\sqrt{\mean{\delta\Delta_{s}^2}}}{\eta_s^{(2,6)}} \sqrt{\frac{m_\eff\gamma_\text{eff}}{k_BT_b\gamma_m}} =2.1\times 10^{4} \text{ MHz/}\mu\text{m},
\end{equation}
which is $\approx 4.6\%$ of $g_m^{\max}$.  Thus for experimental parameters specific to Figure \ref{fig:2Beam_Result}, the cooling limit is closer to $\overline{n}^{(2,6)}_{\min}$ = 2800.

\subsection{``Negative'' differential sensing}\label{subse:OC}

It is interesting to consider the consequences of realizing a \emph{negative} differential sensing factor, $\eta_{p}/\eta_{s}<0$.  Eq. \eqref{eq:feedback} implies that in this case electro-optic feedback can be used to  suppress extraneous noise $(G(f_{m})<0)$ without diminishing  sensitivity to intrinsic motion ($(\eta_{p}/\eta_{s})G(f_{m})>0$).  As a remarkable corollary,  Eqs. \eqref{eq:feedbackPSD} and \eqref{eq:opticaldamping} imply that extraneous noise suppression can coincide with \emph{enhanced} optical damping, i.e., $\mu  = (\eta_{p}/\eta_{s})G(f_{m}) > 0$, if $\eta_{p}/\eta_{s}<0$.   Using electro-optic feedback in this fashion to simultaneously enhance back-action while suppressing extraneous noise seems appealing from the standpoint of optical cooling.

\begin{figure}[t!]
\centering
\includegraphics[width=12cm]{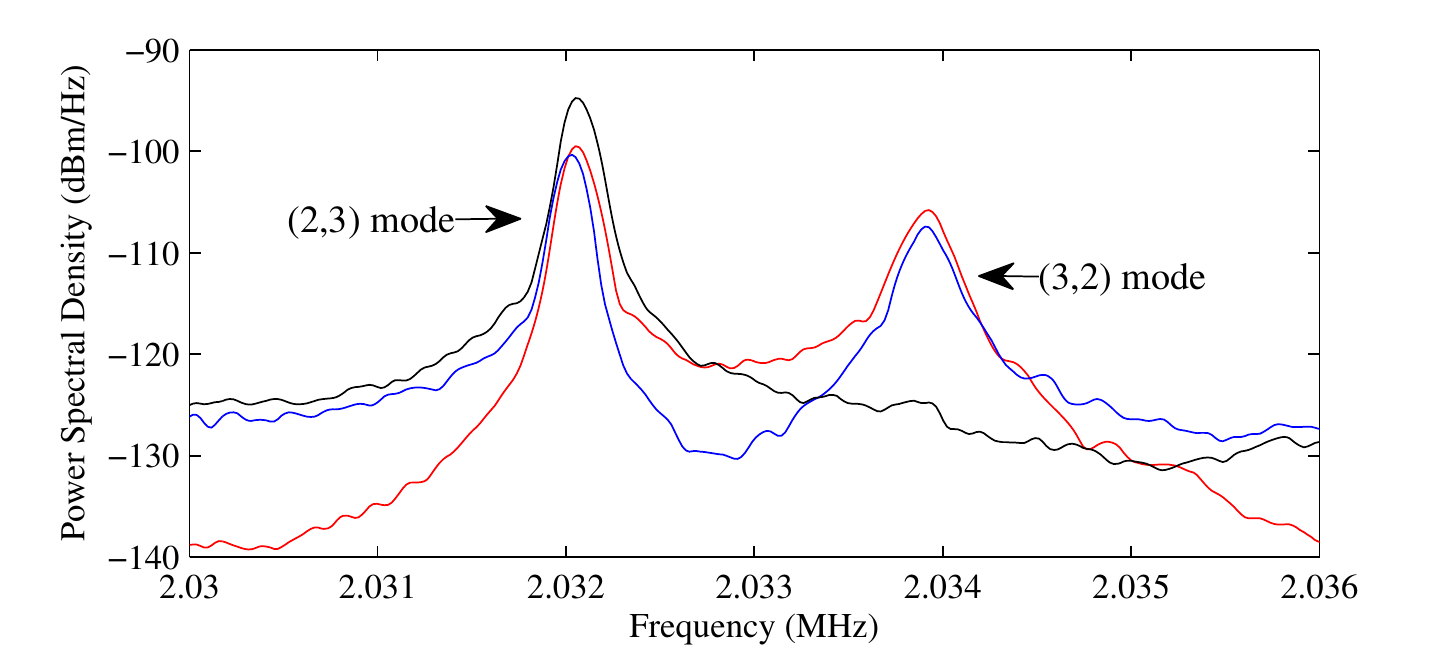}
\caption{Realization of ``negative'' differential sensing for the (3,2) membrane mode.  The TEM$_{00}$ (science) and TEM$_{01}$ (probe) modes are positioned near the center of the membrane.  Measurements of $\delta\Delta_p$ (blue) and $\delta\Delta_s$ (red) are combined electronically on an RF splitter with positive gain, $G_{0}=\eta_{p}^{(3,2)}/\eta_{s}^{(3,2)}$, as discussed in Section \ref{subse:3beam_Exp}.  The power spectrum of the electronic signal before (blue and red) and after (black) the splitter (here in raw units of dBm/Hz) is shown.  The (2,3) noise peak in the combined signal is amplified while the (3,2) noise peak is supressed, indicating that $\eta_{p}^{(2,3)}/\eta_{s}^{(2,3)}>0$ and $\eta_{p}^{(3,2)}/\eta_{s}^{(3,2)}<0$.}
\label{fig:23}
\end{figure}

Achieving $\eta_{p}/\eta_{s}<0$ requires an appropriate choice of cavity and mechanical mode shapes and their relative orientation.  In the context of the ``membrane-in-the-middle'' geometry, we can identify three ways of achieving a negative differential sensing factor.  The first involves selecting a mechanical mode with a nodal spacing comparable to the cavity waist.  Consider the arrangement shown in Figure \ref{fig:OC}, in which the TEM$_{00}$ and TEM$_{01}$ cavity modes are centered on the central antinode of the (1,5) membrane mode (a 1D slice through the midline of the membrane is shown). The ratio of the ($1/e^2$ intensity) cavity waist, $w_{0}$, and the nodal spacing, $w_\text{node}$, has been adjusted so that the distance between antinodes of the TEM$_{01}$ cavity mode roughly matches $w_\text{node}$; in this case, $w_{0}/w_\text{node} = 0.88$.  It is intuitive to see that the displacement profile averaged over the blue (TEM$_{01}$) and red (TEM$_{00}$) intensity profiles is negative in the first case and positive in the second.  Using TEM$_{01}$ as the probe mode and TEM$_{00}$ as the science mode in this case gives $\eta^{(1,5)}_{p}=-0.34$, $\eta^{(1,5)}_{s}=0.37$, and $\eta^{(1,5)}_{p}/\eta^{(1,5)}_{s} = -0.92$ (Appendix \ref{ap:Overlap}).

Another possibility is to center the cavity mode near a vibrational \emph{node} of the membrane. In this situation, numerical calculation shows that rotating the TEM$_{01}$ (probe) mode at an appropriate angle with respect to the membrane can give $\eta_{p}/\eta_{s}<0$ (with TEM$_{00}$ as the science mode), albeit at the cost of reducing the absolute magnitudes of $\eta_{p}$ and $\eta_{s}$.  We have experimentally observed this effect in our system by positioning the cavity modes near the geometric center of the membrane, which is a node for all the modes $(i,j)$ with $i$ or $j$ even.  As in Section \ref{subse:3beam_Exp}, we electronically added and subtracted simultaneous measurements of $\delta\Delta_p$ and $\delta\Delta_s$ in order to assess their differential sensitivity.  A measurement of the noise near the mechanical frequency of the $(2,3)$ membrane mode is shown in Figure \ref{fig:23}. We found for this configuration that \emph{adding} the signals with the appropriate gain (black trace) leads to enhancement of the (2,3) mode (left peak) and suppression of the (3,2) mode (right peak), in contrast to the results in Figure \ref{fig:3Beam_Result}, for which all modes are either suppressed or enhanced.  This suggests that $\eta^{(3,2)}_{p}/\eta^{(3,2)}_{s}<0$.

A third option involves coupling \textit{two} probe fields (with overlap factors $\eta_{p_1,p_2}^{\vec{\sigma}}$ for a membrane mode $\vec{\sigma}$) to  different cavity spatial modes. As in Section \ref{subse:3beam_Exp},  measurements of detuning fluctuations $\delta \Delta_{P1,P2}$ can be electronically combined with differential gain $G_{0}$ to produce a combined feedback signal $\Delta_\cmb(f) = G_0(f)\delta\Delta_{p_1}^{\vec{\sigma}}(f) + \delta\Delta_{p_2}^{\vec{\sigma}}(f)$, which can be mapped onto the science field with gain $G(f) = -(1+G_{0}(f))^{-1}$ (Eq. \eqref{eq:feedback} with $\delta \Delta_{p}$ replaced by $\Delta_{\cmb}$) in order to suppress extraneous noise. A straightforward calculation reveals that the back-action is simultaneously enhanced if $(G_0(f)\eta_{p_1}^{\vec{\sigma}}+\eta_{p_2}^{\vec{\sigma}}) /[(1+G_0(f))\eta_s^{\vec{\sigma}}]<0$. Note that this scheme may require greater overall input power as multiple probe fields are used.

\section{Summary and conclusions}\label{se:Con}

We have proposed and experimentally demonstrated a technique to suppress extraneous thermal noise in a cavity optomechanical system.   Our technique (Section \ref{se:Solution}) involves mapping a measurement of the extraneous noise onto the frequency of the incident laser field, delayed and amplified so as to stabilize the associated laser-cavity detuning.  To obtain an independent measurement of the extraneous noise, we have proposed monitoring  the resonance frequency of an auxiliary cavity mode with reduced sensitivity to the intracavity oscillator but similar sensitivity to the extraneous thermal motion.  %As an important application, we discussed how the technique could be used to ``purify" a field used to optically cool the intracavity oscillator.   We argue that residual feedback of the membrane's motion need not significantly limit, and might in some cases enhance, the optical damping afforded by intrinsic cavity back-action.

To demonstrate the viability of this strategy, we have applied it to an experimental system consisting of a nanomechanical membrane coupled to a short Fabry-P\`{e}rot cavity (Sections \ref{se:Noise} and \ref{se:Exp}).  We have shown that in this system, operating at \emph{room temperature}, thermal motion of the end-mirror substrates can give rise to large laser-cavity detuning noise (Fig. \ref{fig:Sub_Mod}).  Using the above technique, with primary (``science'') and auxiliary (``probe'') cavity modes corresponding to TEM$_{00}$ and TEM$_{01}$, we have been able to reduce this ``substrate'' detuning noise by more than an order of magnitude (Figs. \ref{fig:Exp_Bare_Cavity} and \ref{fig:2Beam_Result}).   We've also investigated how this noise suppression scheme can be used to ``purify'' a red-detuned field used to optically damp the membrane (Section \ref{se:Cooling}).  We found that optical damping is effected by residual coupling of the auxiliary cavity mode to the membrane, producing feedback which modifies the intrinsic cavity ``back-action'' (Fig. \ref{fig:FeedbackRatio}).  We argued that this effect is akin to ``cold-damping'' \cite{OmFbExp:PinardCool99}, and that it need not significantly limit, and could even enhance, the optical cooling, if an appropriate auxiliary cavity mode is used (Section \ref{subse:OC}).  Current challenges include increasing the shot noise sensitivity of the auxiliary probe measurement and reducing or changing the ``sign'' (Section \ref{subse:OC}) of the coupling between the auxiliary cavity mode and the membrane.  

It is worth noting that our technique is applicable to a broader class of extraneous fluctuations that manifest themselves in the laser-cavity detuning, including laser phase noise, radial oscillation of optical fibers, and seismic/acoustic vibration of the cavity structure.  The concept of ``differential sensitivity'' (Eq. \ref{eq:Assumption_x12}) central to our technique is also applicable to a wide variety of optomechanical geometries.  

We gratefully acknowledge P. Rabl from whom we learned the analytic tools for the formulation of our feedback scheme and J. Ye and P. Zoller who contributed many insights, including the transverse-mode scheme for differential sensing implemented in this paper. This work was supported by the DARPA ORCHID program, by the DoD NSSEFF program, by NSF Grant PHY-0652914, and by the Institute for Quantum Information and Matter, an NSF Physics Frontiers Center with support from the Gordon and Betty Moore Foundation.  Yi Zhao gratefully acknowledges support from NSERC.

\begin{appendix}

\section{Opto-mechanical coupling and effective displacement}\label{ap:gandx}

An optical cavity with a vibrating boundary exhibits a fluctuating cavity resonance frequency, $\nu_c (t)=\langle \nu_c \rangle+\delta\nu_c(t)$.   The magnitude of $\delta\nu_c(t)$ depends in detail on the spatial distribution of the intra-cavity field relative to the deformed boundary \cite{OmTheory:Overlap,OmTheory:EffMassPinard}.  It is useful to assign the vibrating boundary a scalar ``effective displacement'' amplitude $\delta x$ and its associated opto-mechanical coupling $g = \delta\nu/\delta x$.  The definition of $g$ and $\delta x$ both depend on the geometry of the system.

We model our system as a near-planar Fabry-P\'{e}rot resonator bisected by a thin dielectric membrane with thickness $\ll \lambda$, as pioneered in \cite{OmExp:HarrisDispersive08}.  In this system,  the electric fields on the left and right sides of the membrane are described by Hermite-Gauss functions $\vec{\psi}(x,y,z)e^{2\pi i\nu t}$ \cite{OmExp:HarrisStrongCoupling10},  here expressed in Cartesian coordinates with the $x$-axis coinciding with the cavity axis, and in units such that $|\vec{\psi}(x,y,z)|^{2}\equiv\vec{\psi}(x,y,z)\cdot\vec{\psi}^{*}(x,y,z)$ is the intensity of the cavity mode at position $(x,y,z)$. Vibrations of mirror 1, mirror 2, and the membrane are each described by a displacement vector field $\vec{u}(x,y,z,t)$. Adapting the treatment in \cite{OmTheory:Overlap}, we write

\begin{equation}\begin{split}
\delta\nu_c(t) \approx &\frac{\partial\nu_c}{\partial \bar{x}_1}\cdot\frac{\int_{A_1}\hat{x}\cdot\vec{u}_1(x,y,z,t) |\vec{\psi}(x,y,z)|^{2}\mathrm{d}A}{\int_{A_1}|\vec{\psi}(x,y,z)|^{2}\mathrm{d}A}\big{\vert}_{\bar{x}_{1}}\\
+&\frac{\partial\nu_c}{\partial \bar{x}_2}\cdot\frac{\int_{A_2}\hat{x}\cdot\vec{u}_2(x,y,z,t) |\vec{\psi}(x,y,z)|^{2}\mathrm{d}A}{\int_{A_2}|\vec{\psi}(x,y,z)|^{2}\mathrm{d}A}\big{\vert}_{\bar{x}_{2}}\\
+&\frac{\partial\nu_c}{\partial \bar{x}_m}\cdot\frac{\int_{A_m}\hat{x}\cdot\vec{u}_m(x,y,z,t) |\vec{\psi}(x,y,z)|^{2}\mathrm{d}A}{\int_{A_m}|\vec{\psi}(x,y,z)|^{2}\mathrm{d}A}\big{\vert}_{\bar{x}_{m}},
\end{split}\label{eq:A1}\end{equation}
where $\bar{x}_{1,2,m}$ are the equilibrium positions of mirror surface 1 ($A_1$), mirror surface 2 ($A_2$), and the membrane ($A_{3}$) at the cavity axis $(y=z=0)$ and $\int_{A_{1,2,m}}$ represents an integral over reflective surfaces $A_{1,2,m}$. $A$ is a plane for the membrane and a spherical sector for each mirror. 

Formula \eqref{eq:A1} can be written in a simplified form
\begin{equation}\begin{split}
\delta\nu_c(t)=& g_{1} \delta x_{1}(t)+g_{2} \delta x_{2}(t)+g_{m}\delta x_{m}(t),
\end{split}\label{eq:A2}\end{equation}
where,
\begin{equation}
g_{1,2,m}\equiv \frac{\partial\nu_c}{\partial \bar{x}_{1,2,m}},
\label{eq:g}
\end{equation} 
represents the frequency shift per unit axial displacement of the equilibrium position and
\begin{equation}
\delta x_{1,2,m}\equiv\frac{\int_{A_{1,2,m}}\hat{x}\cdot\vec{u}_{1,2,m}(x,y,z,t)|\vec{\psi}(x,y,z)|^{2} \mathrm{d}A}{\int_{A_{1,2,m}}|\vec{\psi}(x,y,z)|^{2}\mathrm{d}A}\big{\vert}_{x_{1,2,m}}
\label{eq:apparentdisplacement}\end{equation}
is given by averaging the displacement profile of each surface over the local intra-cavity intensity profile. 

In general, $\delta x_{1,2,m}$ and $g_{1,2,m}$ are functions of $\bar{x}_{1,2,m}$.   Variations in $\delta x_{1,2,m}$ are due to the small change in the transverse intensity profile along the cavity axis.  Expressions for $g_{1,2,m}$ can be obtained by solving explicitly for the cavity resonance frequency as a function of the membrane position within the cavity.  This has been done for a planar cavity geometry in \cite{OmExp:HarrisDispersive08}.  When the membrane (with thickness much smaller than the wavelength of the intracavity field, $\langle\lambda_{c}\rangle$) is placed near the midpoint between the two mirrors, $g_{m}$ is given by \cite{OmExp:HarrisDispersive08}
\begin{equation}
g_{m}=2|r_{m}|g_{0}\frac{\sin(4\pi \bar{x}_{m}/\langle \lambda \rangle)}{\sqrt{1-r_{m}^{2}\cos^{2}(4\pi \bar{x}_{m}/\langle \lambda \rangle)}},
\label{eq:gmembrane}\end{equation} 
where $g_{0} = \langle \nu_c \rangle/L=c/L\langle\lambda_c\rangle$, $r_{m}$ is the reflectivity of the membrane, and the origin of $x$ has been chosen so that $g_{m} = 2r_{m}g_{0}$ when the membrane is positioned halfway between a node and an anti-node of the intra-cavity field (i.e., $\bar{x}_{m} = 3\langle\lambda_c\rangle/8$).  

It can be numerically shown that
\begin{equation}\begin{split}
g_{1} &= g_{0}-\frac{g_{m}}{2},\\
g_{2} &= -g_{0}-\frac{g_{m}}{2},
\end{split}\label{eq:gendmirrors}\end{equation}
which reduces to the expression for a normal Fabry-P\'{e}rot cavity, $g_{1}=-g_{2}=g_{0}$, as $r_{m}\rightarrow0$. 

\section{``Spatial overlap'' factor}\label{ap:Overlap}
Each displacement vector field $u(x,y,z,t)$ can be expressed as a sum over the vibrational eigenmodes of the system:
\begin{equation}
\vec{u}(x,y,z,t)=\sum_{\vec{\sigma}} b^{\vec{\sigma}}(t)\vec{\phi}^{\vec{\sigma}}(x,y,z),
\end{equation}
where $\vec{\phi}_{\vec{\sigma}}$ is the unitless mode-shape function of vibrational eigen-mode $\vec{\sigma}$ and $b_{\vec{\sigma}}$ is a generalized amplitude with units of length.  Effective displacement can thus be expressed:
\begin{equation}
\delta x=\sum_{\vec{\sigma}}\eta^{\vec{\sigma}}b^{\vec{\sigma}}(t),
\label{eq:effectivedispexpansion}\end{equation}
where 
\begin{equation}
\eta^{\vec{\sigma}}\equiv \frac{\int_{A}\hat{x}\cdot \vec{\phi}^{\vec{\sigma}}(x,y,z)|\vec{\psi}(x,y,z)|^{2}\mathrm{d}A}{\int_{A}|\vec{\psi}(x,y,z)|^{2}\mathrm{d}A}
\label{eq:overlap}\end{equation}
is the displacement of vibrational mode $\vec{\sigma}$ averaged over the intra-cavity intensity profile evaluated at surface $A$.  We have referred to $\eta$ as the ``spatial overlap'' factor in the main text.

Vibrational modes $\vec{\sigma} = (i,j)$ of a square membrane (shown, e.g., in Fig. \ref{fig:BeamLocation})  are described by
\begin{equation}
\vec{\phi}^{(i,j)}(\bar{x}_{m},y,z) = \sin\left(\frac{i\pi (y-y_{0})}{d_{m}}\right)\sin\left(\frac{j\pi (z-z_{0})}{d_{m}}\right)\hat{x},
\label{eq:membranemode}\end{equation}
where $d_{m} = 500\;\mu$m is the membrane width and $(y_0,z_0)$ is the offset position of the membrane relative to the cavity axis, located at $(y_0,z_0)=0$.

The transverse intensity profile of the TEM$_{ij}$ cavity mode at planar surface $A_{m}$ coinciding with $x=\bar{x}_{m}$ is given by
\begin{equation}
|\vec{\psi}_{(n,n')}(\bar{x}_{m},y,z)|^{2} = N_{(n,n')}\left(H_{n}\left[\frac{\sqrt{2}\;y}{w(\bar{x}_{m})}\right]H_{n'}\left[\frac{\sqrt{2}\;z)}{w(\bar{x}_{m})}\right]\right)^{2},
\label{eq:TEM}\end{equation}
where $N_{(n,n')}$ is a normalization constant, $H_{n}$ and $H_{n'}$ are the Hermite polynomials of order $n,n'$, respectively, and $w(\bar{x}_{m})\simeq w_{0}\simeq 33.2\;\mu$m is the Gaussian mode radius at position $\bar{x}_{m}$.

Observe that for $y_{0} = z_{0}=0$, all even-ordered membrane modes ($i$ or $j$ = 2,4,6,...) have vanishing spatial overlap factors ($\eta^{(i,j)}=0$), justifying the reduced mode density in the model shown in Figure \ref{fig:Sub_Mem_Mod}.

\section{Multi-mode Thermal Noise}\label{ap:Thermal}

Driven motion of a single vibrational eigen-mode with mode-shape $\vec{\phi}$, amplitude $b$, eigen-frequency $f_{m}$, and frequency-dependent mechanical dissipation $\gamma_{m}(f) \equiv (f_{m}/( Q_{m}))\times(f_{m}/f)$ (here we assume a ``structural damping'' model relevant to bulk elastic resonators \cite{Thermal:Saulson90}) can be described by the linear transformation:
\begin{equation}
(2\pi)^2\left( f_{m}^2-f^2+i f \gamma_{m}(f)\right)m_\eff b(f) \equiv \chi_{m}(f)^{-1}b(f) = F_\text{ext}(f),
\label{eq:mechanicaleqofmotion}\end{equation}
where $F_\text{ext}(f)$ denotes the Fourier transform of the external driving force, $\chi_{m}(f)$ is the mechanical susceptibility, and $m_\eff$ is the effective mass of $b$, defined as
\begin{equation}
m_\eff = \frac{U}{\frac{1}{2}(2\pi f_{m})^2 b^{2}} = \int_{V}\vec{\phi}\cdot\vec{\phi}\;\rho\;dV,
\label{eq:effmass}\end{equation}
where $\rho$ is the mass density of the material.  

Thermal noise is characterized by a random force whose power spectral density \cite{PSD} is given by the Fluctuation-Dissipation Theorem \cite{Thermal:Saulson90,Thermal:IrreversibilityNoise}:
\begin{equation}
S_{F}(f) = \frac{4k_{B}T_b}{2\pi f_{m}}\text{ Im}[\chi_{m}(f)^{-1}]=4k_{B}T_{b}\gamma_{m}(f)m_{\eff}.
\end{equation}

Thermal fluctuations of $b$ are thus given by
\begin{equation}
S_{b}(f)= |\chi_{m}(f)|^2S_{F}(f)=\frac{k_{B}T_b\gamma_{m}(f)}{2\pi^3m_\eff} \frac{1}{(f_{m}^{2}-f^{2})^{2}+f^{2}\gamma_{m}^{2}(f)}.
\label{eq:frequencythermalnoise}\end{equation} 

Using Eq. \eqref{eq:effectivedispexpansion}, the power spectral density of effective displacement associated with $u(x,y,z,t)$ is given by 

\begin{equation}
S_{x}(f)=\sum_{\vec{\sigma}}(\eta^{\vec{\sigma}})^{2}{S_{b^{\vec{\sigma}}}}(f).
\label{eq:apparentthermalnoisespectrum}\end{equation}

Combining Eqs. \eqref{eq:A2} and \eqref{eq:apparentthermalnoisespectrum}, we can express the spectrum of cavity resonance frequency fluctuations for the ``membrane-in-the-middle'' cavity (ignoring radiation pressure effects) as 
\begin{equation}
S_{\nu_{c}}(f) = g_{1}^{2}S_{x_{1}}(f)+g_{2}^{2}S_{x_{2}}(f)+g_{m}^{2}S_{x_{m}}(f).
\label{eq:frequencythermalnoisespectrum}\end{equation}
%where
%\begin{equation}
%S_{x_{1,2,m}}=\sum_{\vec{\sigma}}(\eta^{\vec{\sigma}})^{2}{S_{b_{1,2,m}^{\vec{\sigma}}}}
%\label{eq:apparentthermalnoisespectrum}\end{equation}
%is the one-sided power spectral density of apparent thermal motion \cite{PSD}.

\subsection{Substrate Thermal Noise}\label{ap:SubThermal}

Substrate thermal noise is described by the first two terms in Eq. \eqref{eq:frequencythermalnoisespectrum}.  The main task in modeling this noise is to determine shape and frequency of the substrate vibrational modes.   In our experiment, each end-mirror substrate is a cylindrical block of BK7 glass with dimensions $L_{s}$ (length) $\simeq$ 4 mm, $D_{s}$ (diameter) $\simeq$ 3 mm, and physical mass $m_{p}\simeq$ 60 mg.  The polished end of each substrate is chamfered at 45 degrees, resulting in a reflective mirror surface with a diameter of 1 mm.   The polished face also has a radius of curvature of $R_{s}\simeq$ 5 cm, resulting in a Gaussian mode waist of $w_{0}\simeq 33.2\,\mu$m for cavity length $L=0.74\;\mu$m and an operating wavelength of $\lambda=810$ nm.  

To generate the vibrational mode-shapes and eigenfrequencies associated with this geometry, we use the ``Structural Mechanics'' finite element analysis (FEA) module of Comsol 3.5 \cite{Comsol}.  For each $\vec{\phi}$ we then compute $\eta$ and $m_\eff$ according to Eq. \eqref{eq:overlap} and \eqref{eq:effmass}, respectively.  For the model we also: (a) assume that the substrate is a free mass (ignoring adhesive used to secure it to a glass v-block), (b) ignore the $\approx10\;\mu$m dielectric mirror coating and the radius of curvature on the polished face, (c) assume the following density, Young's modulus, and Poisson's ratio for BK7 glass: $\{\rho,Y,\nu\}=\{2.51\text{ g/cm}^{3}, 81\text{ GPa}, 0.208\}$, (d) assume a TEM$_{00}$ cavity mode centered on the cylindrical axis of the mirror substrates, so that $\eta=0$ (Eq. \eqref{eq:overlap}) for all non-axial-symmetric modes, (e) assume $g_{1}=-g_{2}=g_{0}$ and $g_{m}=0$ in Eq. \eqref{eq:frequencythermalnoisespectrum}.  To obtain good qualitative agreement to the measured data, we have adjusted (via the physical mass) the predicted fundamental vibrational frequency substrate 1 and substrate 2 by factors of 0.995 and 0.998, respectively, from the results obtained from the FEA model.  We have also assumed a uniform quality factor of $Q_{m} = f_{m}/(\gamma_{m}(f_{m}))=700$ for all internal modes in Eq. \eqref{eq:frequencythermalnoise}.

\subsection{Membrane Thermal Noise}\label{ap:MemThermal}

Membrane thermal noise is described by the last term in  Eq. \eqref{eq:frequencythermalnoisespectrum}.  For a square membrane with mode-shape functions as given in \eqref{eq:membranemode} (i.e., $\phi=1$ for an antinode), the effective mass as defined in Eq. \eqref{eq:effmass} is $m_\eff=m_{p}/4$ for all modes.  

In the main text we model membrane motion using a velocity-damping model, i.e., $\gamma_{m}(f)=\gamma_{m}(f_{m})=f_{m}/Q_{m}$.  We use the short-hand notation $\gamma_{m}(f_{m})\equiv\gamma_{m}$.  Taking into account cavity back-action as described in Appendix \ref{ap:suppress_length_frequency}, the thermal noise spectrum of membrane mode amplitude $b$ takes on the approximate form
\begin{equation}
S_{b}(f)\approx \frac{ k_{B}T_{b}\gamma_{m} }{2\pi^3 m_\eff}\frac{1}{((f_m+\delta f_{\opt})^{2}-f^{2})^{2}+f^{2}(\gamma_m+\gamma_{\opt})^2}
\label{eq:membranePSD}\end{equation}
near resonance.  

Expressions for the optical spring shift, $\delta f_\opt$, and damping rate, $\gamma_\opt$, depend on cavity parameters and electro-optic feedback parameters, as detailed in Appendix \ref{ap:suppress_length_frequency}.
Optical damping gives rise to optical cooling as characterized by reduction of the mean-squared vibrational amplitude, $\mean{b^2}$. From Eq. \eqref{eq:membranePSD} we obtain:
\begin{equation}\label{eq:appendixcooling}
\langle b^2 \rangle = \int_{0}^{\infty} S_{b}(f)\mathrm{d}f = \frac{k_{B}T_{b}}{(2\pi f)^2 m_{\eff}}\frac{\gamma_{m}}{\gamma_{m}+\gamma_{\opt}},
\end{equation}
which is duplicated in Section \ref{se:Cooling} as Eq. \eqref{eq:opticalcooling_NoFB}.
%\hl{Note that, the impact of electro-optic feedback on membrane thermal noise is not considered here} (\ref{ap:MemThermal}), \hl{and will be addressed in the following section.}

\section{Thermal noise suppression with electro-optic feedback: optomechanical interaction}\label{ap:suppress_length_frequency}
Here we consider a simple model for the dynamics of a vibrating membrane (or equivalently, a vibrating mirror) linearly coupled to an optical cavity driven by a frequency modulated laser.  We focus on a single vibrational mode of the membrane with amplitude $b(t)$.  The cavity exhibits a fluctuating resonance frequency $\delta\nu_{c}(t) = g_{m}\eta_{s} b(t)$, where $g_{m}$ is the optomechanical coupling (Eq. \ref{eq:g}) and $\eta_{s}$ the spatial overlap factor between the cavity mode and the vibrational mode (Eq. \ref{eq:overlap}).    Fluctuations $\delta\nu_{c}(t)$ give rise to intracavity intensity fluctuations, which alter the dynamics of $b(t)$ through the radiation pressure force.  Superimposed onto this back-action is the effect of electro-optic feedback, which we model by assuming a definite phase relationship between $b(t)$ and the modulated frequency of the incident field.  % and study the effect of this auxiliary feedback on the dynamics of the system.

\subsection{Optomechanical equations of motion} 
We adopt the following coupled differential equations to describe vibrational amplitude $b(t)$ and intracavity field amplitude $a(t)$ (here expressed in the frame rotating at the frequency of the drive field, $\langle \nu_{0}\rangle = \langle \nu_{c}\rangle+\langle \Delta \rangle$, and normalized so that $|a(t)|^2=U_{c}(t)$ is the intracavity energy):
\begin{subequations}\label{eq:apDE}\begin{align}
\ddot{b}(t)+\gamma_{m}\dot{b}(t)+(2\pi f_{m})^2 b(t)&=F_\text{ext}(t)+\delta F_{\rad}(t),\\
\dot{a}(t)+2\pi(\kappa+i(\langle \Delta \rangle-\delta\nu_c(t)))a(t) &=\sqrt{4\pi\kappa_1}E_0e^{2\pi i\phi(t)}.
\end{align}\end{subequations}

Eq. (\ref{eq:apDE}a) describes the motion of a velocity-damped harmonic oscillator driven by an external force $F_\text{ext}(t)$ in addition to a radiation pressure force $F_{\rad}(t)=\langle F_{\rad}\rangle +\delta F_{\rad}(t)$.  We define $b$ relative to its equilibrium position with the cavity field excited, thus we ignore the static part of the radiation pressure force.  We adopt the following expression for the radiation pressure force based on energy conservation: $F_{\rad} = -\partial U_{c}/\partial b= -g_{m}\eta U_{c}/\langle \nu_{c} \rangle$.

 Eq. (\ref{eq:apDE}b) is based on the standard input-output model for a low-loss, two-mirror resonator \cite{CavityInOut}.  Here $E_{0}$, $|E_{0}|^2$ and $\phi(t)$ represent the amplitude, power and instantaneous phase of the incident field, respectively, and $\kappa = \kappa_{1}+\kappa_{2}+\kappa_{L}$ the total cavity (amplitude) decay rate, expressed as the sum of the decay rates through M$_{1,2}$ and due to internal losses, respectively.  Note that in the ``membrane-in-the-middle'' system, $\kappa_{1,2,L}$ all depend on the membrane's position, $\bar{x}_{m}$. The instantaneous frequency and detuning fluctuations of the input field are given by:
\begin{subequations}\begin{align}
\delta\nu_{0}(t) &= \dot{\phi}(t)\\
\delta\Delta(t) &= \dot{\phi}(t)-\delta\nu_{c}(t)= \dot{\phi}(t)-g_{m}\eta_{s}b(t).
\end{align}\label{eq:dphi}\end{subequations}

We seek solutions to Eq. \eqref{eq:apDE} by expressing $a(t)$ as a small fluctuation around its steady state value, i.e., $a(t)=\langle a \rangle +\delta a(t)$.  In this case the radiation pressure force is given by 
\begin{equation}
%b(t) &=\langle b \rangle+b(t)\\
%\delta a(t) &= a(t)-\langle a \rangle\\
\delta F_{\rad}(t) = \frac{g_{m}\eta}{\nu_{c}}(\langle a \rangle^{*} \delta a(t)+\langle a \rangle\delta a^{*}(t)).
\label{eq:linearizedequations}\end{equation}
An equation of motion for $\delta a(t)$ is obtained from Eq. \eqref{eq:apDE} with the assumption that $\delta a(t)\ll \langle a \rangle$, $\phi(t)\ll 1$, and $\delta\nu_c(t)\ll\langle \Delta \rangle$.  This approximation gives
\begin{equation}
%\ddot{b}(t)+\gamma_{m}\dot{b}(t)+(2\pi f_{m})^2 b(t)&=F_\text{ext}(t)-\frac{g_{m}\eta}{\langle \nu_{c}\rangle}\delta |a(t)|^2\\
\dot{\delta a}(t)+2\pi(\kappa+i(\langle \Delta \rangle)\delta a(t)-2\pi\kappa\delta\nu_{c}(t) \langle a\rangle=2\pi i\sqrt{4\pi\kappa_1}E_0\phi(t),
\label{eq:fieldeqnfirstorder}\end{equation}
where
\begin{equation}
\langle a \rangle = E_{0}\sqrt{4\pi\kappa_{1}}/(\kappa+i\langle\Delta\rangle).
\label{eq:asteady}\end{equation}

Applying the Fourier transform \cite{PSD} to both sides of Eq. (\ref{eq:fieldeqnfirstorder}), we obtain the following expression for the spectrum of intracavity amplitude fluctuations:
\begin{equation}
\delta a(f) = \frac{i\langle a \rangle}{\kappa+i(\langle \Delta \rangle+f)}\left(\delta\nu_{c}(f)+(\kappa+i\langle \Delta \rangle)\phi(f)\right).
\end{equation}

Combining expressions for $\delta a(f)$ and $[\delta a^{*}](f)$ (the Fourier transform of $\delta a^{*}(t)$, which obeys the complex conjugate of Eq. \eqref{eq:fieldeqnfirstorder}), we obtain the following expression for the spectrum of radiation pressure force fluctuations from the Fourier transform of Eq. \eqref{eq:linearizedequations}:
\begin{equation}
\delta F_{\rad}(f) = \frac{g_{m}\eta|\langle a \rangle|^2}{\nu_{c}}\left(\frac{i}{\kappa+i(\langle \Delta \rangle+f)}-\frac{i}{\kappa+i(\langle \Delta \rangle+f)}\right)(\delta\nu_{c}(f)-i\phi(f)/f).
\end{equation}

Using Eq. \eqref{eq:asteady} and identifying $\delta\nu_{c}(f)-i\phi(f)/f = -\delta\Delta(f)$ as the instantaneous detuning fluctuations, we infer the following relationship for the radiation pressure force fluctuations:
\begin{subequations}\begin{align}
\delta F_{\rad}(f) &= \frac{g_{m}\eta}{\langle \nu_{c}\rangle}|E_{0}|^2\frac{4\pi \kappa_{1}}{\kappa^2+\langle \Delta \rangle^2}\left(\frac{i}{\kappa+i(\langle \Delta \rangle+f)}-\frac{i}{\kappa+i(\langle \Delta \rangle+f)}\right)\delta\Delta(f)\\
&\equiv\varphi(f)\delta\Delta(f) =  -\varphi(f) g_{m}\eta_{s} b(f)(1+\mu(f)),
\end{align}\label{eq:radiationpressurewithfeedback}\end{subequations}
where $\mu(f)=-(\eta_{p}/\eta_{s})G(f)$ is the electro-optic feedback gain as defined in Eq. \eqref{eq:feedback}.

Eq. \eqref{eq:radiationpressurewithfeedback} suggests that radiation pressure force fluctuations associated with thermal motion can be suppressed by using electro-optic feedback to stabilize the associated laser-cavity detuning fluctuations.  The effect of $\delta F_{\rad}$ on the dynamics of $b$ in Eq. (\ref{eq:fieldeqnfirstorder}a) may be expressed as a modified mechanical susceptibility, $\chi_{\eff}(f) \equiv b(f)/F_\text{ext}(f)$.  Applying the Fourier transform to both sides of Eq. (\ref{eq:apDE}a) and using (\ref{eq:radiationpressurewithfeedback}) gives:
\begin{subequations}\begin{align}
\chi_{\eff}(f)^{-1} &= \chi_{m}(f)^{-1}-\delta F_{\rad}(f)/b(f)\\
&=(2\pi)^2\left(f_{m}^2-f^2+i f \gamma_{m}\right)m_\eff +\varphi(f) g_{m}\eta_{s}(1+\mu(f)).
\end{align}\end{subequations}

 For sufficiently weak radiation pressure, the mechanical susceptibility near resonance is
\begin{equation}
\chi_{\eff}(f)^{-1} \approx(2\pi)^2\left((f_{m}+\delta f_{\opt})^2-f^2+i f (\gamma_{m}+\gamma_{\opt})\right)m_\eff%+\varphi(f_{m}) g_{m}\eta_{s}(1+\mu(f_{m}))
\end{equation}
where
\begin{subequations}\begin{align}
\gamma_{\opt}&\equiv -\frac{1}{(2\pi)^{2}}\frac{1}{f_{m}m_{\eff}}\text{Im}\left(\frac{F_{\rad}(f_{m})}{b(f_{m})}\right)= \frac{1}{(2\pi)^{2}}\frac{g_{m}\eta_{s}}{f_{m}m_{\eff}} \text{Im}\left((1+\mu(f_{m}))\varphi(f_{m})\right),\\
\delta f_{\opt}&\equiv -\frac{1}{(2\pi)^{2}}\frac{1}{2f_{m}m_{\eff}}\text{Re}\left(\frac{F_{\rad}(f_{m})}{b(f_{m})}\right)= \frac{1}{(2\pi)^{2}}\frac{g_{m}\eta_{s}}{2f_{m}m_{\eff}} \text{Re}\left((1+\mu(f_{m}))\varphi(f_{m})\right).
\end{align}\label{eq:dampingandspring}\end{subequations}

From this expression and the Fluctuation-Dissipation theorem (Eq. \eqref{eq:membranePSD}) we infer the modified membrane thermal noise spectrum used in Appendix \ref{ap:MemThermal}:
\begin{equation}
S_{b}(f) = |\chi_{\eff}(f)|^2\cdot 4k_{B}T_b\gamma_m f= \frac{ k_{B}T_{b}\gamma_{m} }{2\pi^3 m_\eff}\frac{1}{((f_{m}+\delta f_{\opt})^{2}-f^{2})^{2}+f^{2}(\gamma_m+\gamma_{\opt})^2}.
\end{equation}

 %for the modified susceptibility $\chi_{eff}(f)$ we infer that following modified damping rate and frequency of the oscillator:
%\begin{subequations}\begin{align}
%\gamma_{eff}\equiv\frac{\text{ Im}[\chi_{eff}(f_{m})^{-1}]}{(2\pi)^2m_{eff}f_{m}}=\gamma_{m}+\gamma_{opt}(f) 
%=\text{ Im}[\chi_{eff}(f)^{-1}]
%&= (2\pi)^2\left(f_{m}^2-f^2+i f \gamma_{m}\right)m_\eff -\varphi(f) g_{m}\eta_{s}(1+\mu(f))
%\end{align}\end{subequations}
\end{appendix}


\begin{thebibliography}{10}
\bibitem{PSD}
For a generic time-dependent variable $\zeta(t)$, the Fourier transform is
  defined as $\zeta(f)\equiv \int_{-\infty}^\infty \zeta(t)e^{-2\pi i f t}
  \mathrm{d}t$, and the one-sided power spectral density at Fourier frequency
  $f$ is $S_\zeta(f) \equiv
  2\int_{-\infty}^\infty\mean{\zeta(t)\zeta(t+\tau)}e^{2i\pi ft}\mathrm{d}\tau$
  with unit $[\zeta]^2/$Hz. The normalization convention is that
  $\mean{\zeta^2(t)}=\int_0^\infty S_\zeta(f)\mathrm{d}f$.
\end{thebibliography}
\end{document}